\documentclass[journal]{IEEEtran}
\usepackage{cite}
\usepackage{amsmath,amssymb,amsfonts}
\usepackage[noend]{algpseudocode}

\algnewcommand\And{\textbf{and}}
\algnewcommand\Or{\textbf{or}}
\usepackage{graphicx}
\usepackage{textcomp}
\usepackage{xcolor}
\graphicspath{{./figures/}}
\usepackage{flushend}
\usepackage{accents}

\usepackage{calc}
\usepackage{slashbox}
\usepackage{graphics}
\usepackage{latexsym}
\usepackage[dvips]{epsfig}
\usepackage{tabularx}
\usepackage{arydshln}
\usepackage{subcaption}
\usepackage[ruled,vlined,linesnumbered]{algorithm2e}
\usepackage[noend]{algpseudocode}
\usepackage[linewidth=1pt]{mdframed}

\allowdisplaybreaks[1]

\usepackage[T1]{fontenc}
\usepackage{inconsolata}

\title{Low-Latency Network-Adaptive Error Control for Interactive~Streaming}

\author{Salma~Emara,~\IEEEmembership{Member,~IEEE,} Silas~L.~Fong, ~\IEEEmembership{Member,~IEEE,} Baochun Li,~\IEEEmembership{Fellow,~IEEE,} Ashish Khisti, ~\IEEEmembership{Member,~IEEE,} Wai-Tian Tan, ~\IEEEmembership{Senior Member,~IEEE,} Xiaoqing~Zhu,~\IEEEmembership{Senior Member,~IEEE,} and John~Apostolopoulos,~\IEEEmembership{Fellow,~IEEE}%
\thanks{Manuscript created and submitted on May 1, 2020.}
\thanks{This paper was presented in part at 27th ACM International Conference on Multimedia, Nice, France, 2019.}
\thanks{S.~L.~Fong is with Qualcomm Flarion Technologies, NJ 08807, USA (\text{E-mail:} \texttt{silas.fong@ieee.org}).}
\thanks{S.~Emara, B.~Li and A.~Khisti are with the Department of Electrical and Computer Engineering, University of Toronto, Toronto, ON M5S 3G4, Canada (E-mails: \texttt{salma@ece.utoronto.ca}, \texttt{bli@ece.utoronto.edu},  \texttt{akhisti@ece.utoronto.ca}).}
\thanks{W.-T.~Tan, X.~Zhu and J.~Apostolopoulos are with Cisco Systems, San Jos\'e, CA 95134, USA.}
}

\begin{document}

\IEEEtitleabstractindextext{%

\begin{abstract}
We introduce a novel network-adaptive algorithm that is suitable for alleviating network packet losses for low-latency interactive communications between a source and a destination. Our network-adaptive algorithm estimates in real-time the best parameters of a recently proposed streaming code that uses forward error correction (FEC) to correct both arbitrary and burst losses, which cause a crackling noise and undesirable jitters, respectively in audio. In particular, the destination estimates appropriate coding parameters based on its observed packet loss pattern and sends them back to the source for updating the underlying code. Besides, a new explicit construction of practical low-latency streaming codes that achieve the optimal tradeoff between the capability of correcting arbitrary losses and the capability of correcting burst losses is provided. Simulation evaluations based on statistical losses and real-world packet loss traces reveal the following: (i) Our proposed network-adaptive algorithm combined with our optimal streaming codes can achieve significantly higher performance compared to uncoded and non-adaptive FEC schemes over UDP (User Datagram Protocol); (ii) Our explicit streaming codes can significantly outperform traditional MDS (maximum-distance separable) streaming schemes when they are used along with our network-adaptive algorithm.
\end{abstract}
}
\maketitle
\IEEEdisplaynontitleabstractindextext
\IEEEpeerreviewmaketitle


\section{Introduction}
\label{sec:introduction}
\IEEEPARstart{S}{everal} current and emerging applications over the Internet demand real-time interactive streaming, such as high-definition video conferencing, augmented/virtual reality and online gaming. At the core of these low-latency applications is the need to deliver packets reliably and with low-latency to provide the user with the expected functionality and responsiveness. Given that the Internet is a packet-switched network, where reliable packet delivery is not guaranteed, the urge for effective methods to protect live video communications over the Internet has never been greater.


At the network layer, packet erasures or losses over the Internet are inevitable. Packets are lost due to unreliable wireless links or congestion at network bottlenecks. To recover missing packets introduced by the network layer, two basic methods at the transport layer have been widely implemented: Automatic repeat request (ARQ) and forward error correction (FEC). 

ARQ is a retransmission-based scheme, where the transmitter retransmits a packet based on feedback from receiver. If the communication is between distant users, the extra round-trip delay incurred by the retransmission maybe intolerable for real-time streaming applications. Correcting an erasure using ARQ yields a 3-way delay (forward + backward + forward). 



On the other hand, FEC does not require any retransmission. Instead FEC schemes increase the correlation among the transmitted symbols by sending redundant information. Using this redundant information, erased packets can be reconstructed using the surviving or correctly received packets. Low-density parity-check (LDPC) codes ~\cite{Gall}\cite{MacKayNeal1997}  and digital fountain codes~\cite{Luby2002}\cite{Shokrollahi2006} are two FEC schemes that are currently used in the DVB-S2~\cite{DVB-S2} and DVB-IPTV~\cite{DVB-IPTV} standards for non-interactive streaming applications. Besides, there is on-going research on improving these codes for applications such as scalable video transmission~\cite{yuan-tmm16}. These codes operate over long block lengths, typically a few thousand symbols \footnote{A symbol is an element in the finite field. For example, codes over GF($2^m$) can have each symbol taking one of $2^m$ values, and is represented as an $m$-bit symbol. A code over GF(256=$2^8$) has a symbol of $1$ byte}. 

Noteworthy caveats of LDPC and fountain codes involve i.) the need to wait for the arrival of longer block lengths and ii.) larger computation time to encode and decode these long block lengths. Thus, LDPC and fountain codes are preferable for applications in which the delay constraints are not stringent. However, when the delay constraints are strict and block lengths are short (e.g. a few hundred symbols) -- like in low-latency streaming applications, LDPC and digital fountain codes become unfit. 


Apart from LDPC and digital fountain codes, for interactive applications, Reed Solomon (RS) and Random Linear Convolutional (RLC) codes can be used to recover losses over shorter block lengths as in~\cite{yang-tmm2005, stankovic-tmm04, hameed-tmm16}. However, the pattern of erasures (i.e. bursty or arbitrary erasures) play a role in the efficiency of a coding scheme. For example, to recover burst losses, longer RS codes are required resulting in higher delay. Meanwhile, RLC codes can have the same correcting capacity of RS codes, but with lower delay to recover bursty erasures than RS codes~\cite{BKTAmagazine17}.  To better illustrate the effect of different loss patterns, bursty losses -- usually induced by network congestion -- lead to crackling noise, and arbitrary erasures -- generated by unreliable wireless links-- result in undesirable jitters/pauses in audio. Therefore, optimal and fast recovery for all loss types is influencing the user experience. 

Interestingly, having a block or convolutional structure by itself does not yield a minimum recovery delay. To achieve lower recovery delay, FEC schemes that reconstruct earlier lost packets without waiting to correct all losses provide the basis for streaming codes (i.e. low-latency FEC schemes). Efforts in designing low-latency FEC schemes that operate over short block lengths to improve interactive communication is observable in Raptor codes~\cite{3GPP-TS-26.346}, RaptorQ codes~\cite{ietf-raptor}, randomized linear codes~\cite{ietf-rlc} and others ~\cite{MIT-CSAIL-TR-2010-031, HSP2013, Shokrollahi2006}, and many works have used these codes in video streaming ~\cite{chen-tmm20, wu-tmm18}.

Surely, the employment of FEC schemes to recover lost voice packets has lead to the success of Skype~\cite{Skype2010}. In addition, adaptive hybrid NACK/FEC has been deployed in WebRTC to acquire a better trade-off between temporal quality, spatial video quality and end-to-end delay~\cite{HSP2013}. However, still FEC is not optimized for low-latency in to recover burst losses.

If we follow existing FEC technologies (e.g., WebRTC~\cite{HSP2013}, Skype~\cite{Skype2010}, Raptor codes~\cite{3GPP-TS-26.346}, RaptorQ codes~\cite{ietf-raptor} and randomized linear codes~\cite{ietf-rlc}) and choose the parity frames based on coding over the past multimedia frames using maximum-distance separable (MDS) codes, the resulting FEC streaming code is optimal for correcting arbitrary losses subject to the decoding delay $d_{\mathrm{d}}$, but not for bursty losses. This is the research gap that we target, i.e. deficiency in an optimal FEC streaming code that can correct both arbitrary and bursty erasures subject to decoding delay $d_{\mathrm{d}}$.

\subsection{Main Contributions}
Extensive research has been conducted to study the abilities of streaming codes in recovering bursty and arbitrary (isolated) losses~\cite{FKLTZA2018, KrishnanKumar2018}. The authors in~\cite{FKLTZA2018} and \cite{KrishnanKumar2018} have {proposed a high-complexity construction of a class of FEC streaming codes that has the following properties:
\begin{enumerate}
\item[(a)] Correct both arbitrary and burst erasures
\item[(b)] Achieve the optimal trade-off between correcting arbitrary and burst erasures under a given maximum delay constraint
   \end{enumerate}
   
This motivates us to \emph{design} a low-latency error control scheme based on low-complexity FEC streaming codes to satisfy Properties (a) and (b) and \emph{implement} the design for employment in real-world networks. 
Our real-time error control design has the following features:
\begin{itemize}
\item[(i)] A new explicit construction of low-latency streaming codes achieving a delay-constraint optimal trade-off between the capability of correcting arbitrary erasures and the capability of correcting burst erasures.
\item[(ii)] A network-adaptive algorithm that updates the parameters of our constructed low-latency streaming codes in real-time to adapt to varying network conditions. The destination estimates appropriate coding parameters based on its observed erasure pattern, and then the estimated parameters are fed back to the source for revising the coding parameters.
   \end{itemize}
   
To evaluate our network-adaptive FEC streaming design, we conduct extensive simulations and real-world experiments. The results of our simulations and real-world experiments show that our network-adaptive scheme perform remarkably better compared to uncoded and non-adaptive FEC schemes over UDP in terms of reliability. We extend our comparisons to adaptive MDS streaming codes, and we show that our network-adaptive code outperforms MDS streaming codes in highlighted scenarios and performs a lot better in terms of latency. 

Note that the maximum delay constraint that appeared in Property (b) and Property (i) is a suitable delay metric for interactive communication, because any packet recovered beyond a certain threshold would yield undesirable pauses. On the contrary, the average delay metric is more relevant for non-interactive streaming applications such as video streaming, which has been used in \cite{JKW2012, JKW2014, MMY2019, CMBM2019} to study the tradeoff between throughput and average delay for non-interactive streaming where no packet is ever discarded.
Besides our primary results in ~\cite{FEBKTCA2019}, we add: 
\begin{enumerate}
\item[(a)] Reproducible simulation results over Fritchman channel.
\item[(b)] Experimental results expressed in terms of comprehensive metrics including frame loss rate (FLR), coding rate and Perceptual Evaluation of Speech Quality (PESQ) score~\cite{ITU_G862.2}.
\item[(c)] A performance comparison between our explicit FEC streaming code along with our network-adaptive algorithm and state-of-the-art FEC streaming codes.
\end{enumerate}

\section{Concept of FEC Streaming Codes}
\label{conceptFEC}

\begin{figure}[t!]
\begin{minipage}[b]{\linewidth}
\centering
\includegraphics[width=.6\linewidth, bb=0 0 200 200]{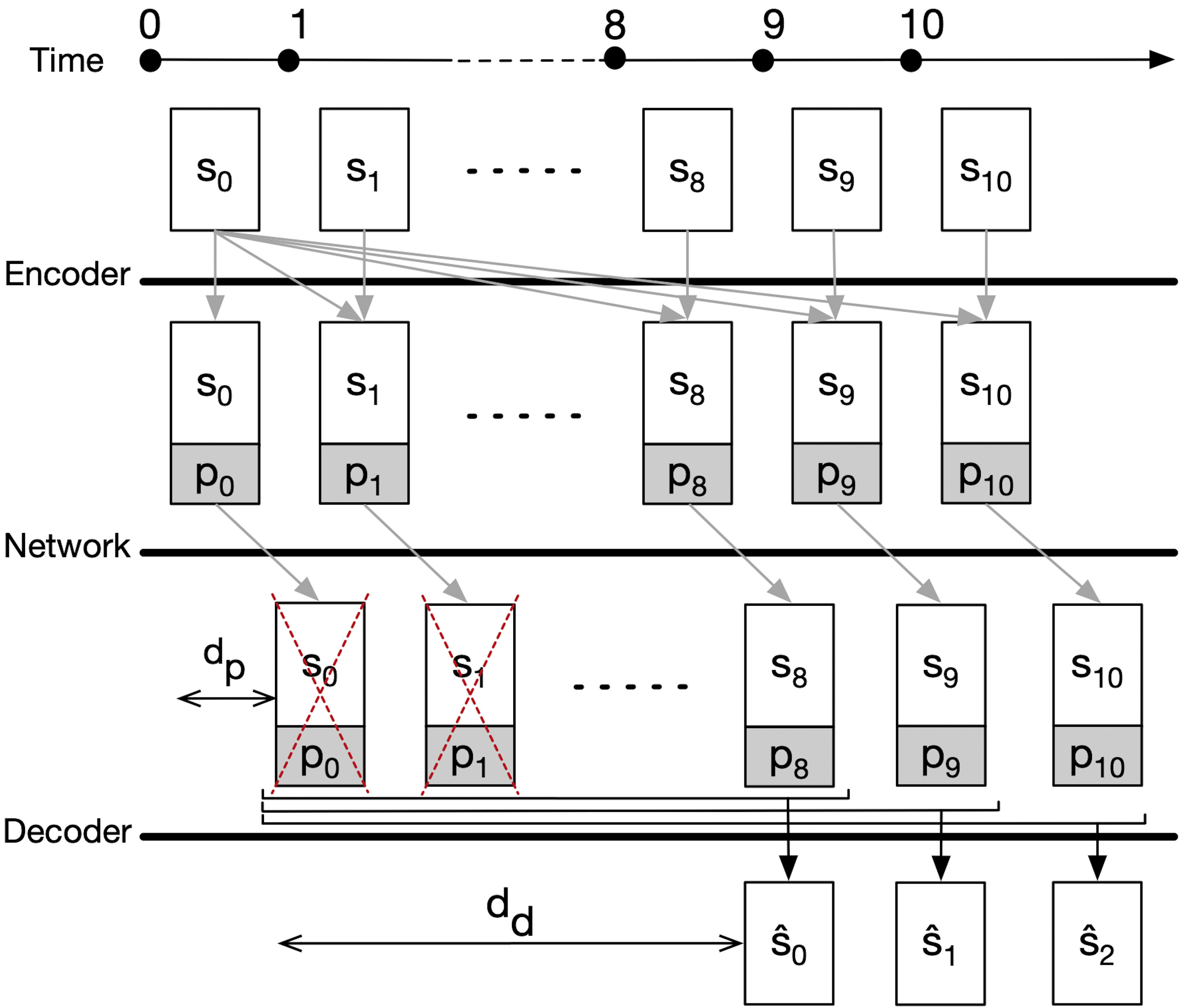}
\caption{The general framework of FEC to recover dropped packets} \label{figure_FEC_concept}
\end{minipage}%
\end{figure}
\begin{figure}[t!]
\begin{minipage}[b]{\linewidth}
\centering
\includegraphics[width=.75\linewidth, bb=0 0 1000 200]{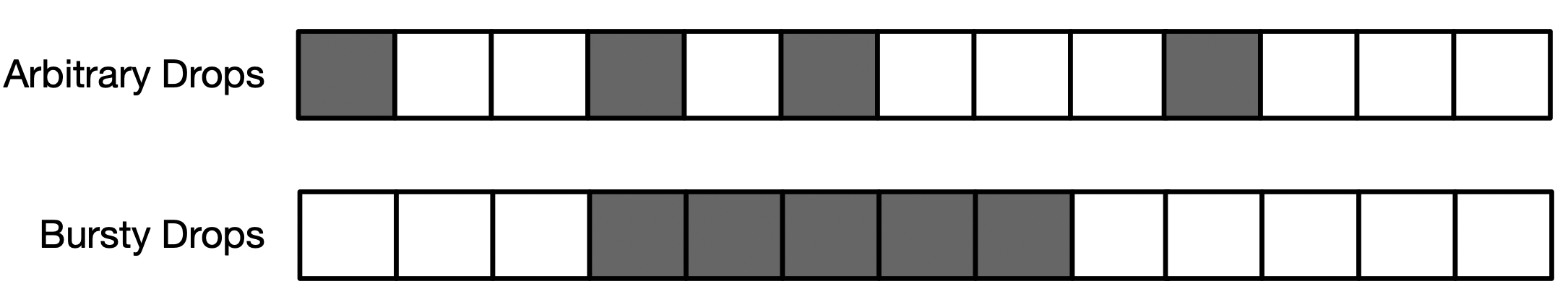}
\caption{Illustration of burst and arbitrary drops, with packet drops marked by dark squares}
\label{figureErasureType}
\end{minipage}%
\end{figure}
The general framework of FEC streaming code is illustrated in Fig.~\ref{figure_FEC_concept}. The source/sender periodically generates a sequence of multimedia frames~$s_i$, where~$i$ is the number of the frame. Each multimedia frame is concatenated with a parity frame~$p_i$, and they are encapsulated in a network packet which travels to the destination with a propagation delay~$d_{\textrm{p}}$. An alternative approach is to transmit parity information as separate packets. Our approach has the advantage of not creating more packets, which lowers overhead for transmission over WiFi, and also simplifies analysis.

The network packets may be dropped by the network. The drops may be in an arbitrary manner due to unreliable (wireless) links or in a bursty manner due to network congestion. The pattern of drops can be arbitrary or bursty in nature, as illustrated in Fig.~\ref{figureErasureType}. The destination aims to recover the multimedia frames sequentially subject to a decoding delay constraint~$d_{\textrm{d}}$, where lost multimedia frames can be recovered with the help of subsequent parity frames. For example, if packets~$0$ and~$1$ are dropped as illustrated in Fig.~\ref{figure_FEC_concept}, then the parity frames in packets~$2$ to $8$ may help recover frame~$0$ with decoding delay of~$8$ frames, and the parity frames in packets~$2$ to $9$ may help recover frame~$1$ with the same decoding delay.

We discuss earlier that if we follow existing technologies such as WebRTC~\cite{HSP2013} and Skype~\cite{Skype2010} in choosing the parity frames based on coding over the past multimedia frames using MDS codes, the resulting FEC streaming code is optimal for correcting arbitrary losses subject to the decoding delay~$d_{\textrm{d}}$, but not bursty losses. In other words, to make it optimal in correcting bursty losses as well, we would require more time leading to an increase in decoding delay~$d_{\textrm{d}}$.
However, in order to achieve the optimal tradeoff between the capability of correcting arbitrary losses and the capability of correcting burst losses subject to a decoding delay constraint, we have to carefully choose the parity frames. The existence of such optimal parity frames has been recently proved in~\cite{FKLTZA2018,KrishnanKumar2018}. More precisely, given a decoding delay constraint of~$T$ frames and any coding parameters~$(B,N)$ such that~$T\ge B\ge N\ge 1$, there exists an optimal choice for the parity frames which leads to a capacity-achieving streaming code that can correct length-$B$ burst erasures and~$N$ arbitrary erasures.


\section{Explicit Construction of Optimal Streaming Codes over GF(256)} 
\label{construction}
References~\cite{FKLTZA2018,KrishnanKumar2018} proves the existence of optimal streaming codes, but their research lacks explicit construction of optimal streaming codes over practical field size.
 \smallskip
\cite{FKLTZA2018,KrishnanKumar2018} state that for any $T\ge B\ge N\ge 1$, the $(T, B, N)$-capacity equals $\mathrm{C}(T,B,N)$ as defined in~\eqref{capacityPTP}. This motivates that for any $T\ge B\ge N\ge 1$, an $(n,k,T)_{\mathbb{F}}$-code that corrects any $(B,N)$-erasure sequence is said to be \emph{optimal} if $\frac{k}{n}=\mathrm{C}(T,B,N)$.

 \begin{align}
\mathrm{C}(T,B,N)\triangleq
\frac{T-N+1}{T-N+B+1}. \label{capacityPTP}
\end{align}

We present the first explicit construction of optimal streaming codes over GF(256) when $T\le 11$. Readers who are uninterested in the explicit construction may refer to the supplementary material of this manuscript or ~\cite{FEBKTCA2019}. Uninterested readers may take the following result for granted:
\begin{quote}``Let $\mathbb{F}=\text{GF(256)}$. For any $T \ge B \ge N \ge 1$, an $(n,k,T)_\mathbb{F}$-code that corrects any $(B,N)$-erasure can be efficiently generated."
\end{quote}


\section{Network-Adaptive Algorithm} \label{algorithm}
\subsection{A Conservative Algorithm that Estimates Channel Parameters $B$ and $N$ in~$L$ Channel Uses}
\begin{algorithm} 
\small
\DontPrintSemicolon
\SetAlgoLined
\SetKwInOut{Input}{Inputs}
\SetKwInOut{Output}{Outputs}
\SetKw{Continue}{continue}
\KwResult{$\hat B_i$ and $\hat N_i$ are generated at the destination for every packet~$i$ where $0\le i\le L-1$. All correctible length-$(T+1)$ erasure patterns that occur by channel use~$i$ can be perfectly recovered by any code that corrects all $(\hat B_i,\hat N_i)$-erasures.}
\Input{~\:$T$, $L$ and $e^L$ denoting decoding delay, duration, and length-$L$ erasure pattern respectively.}
\Output{~\:$\hat B_i$ and~$\hat N_i$ for every packet~$i$.}
\BlankLine
$\mathrm{previous\_seq\_\#} \gets -1$\;
$(\hat B_{-1},\hat N_{-1}, N_{\max}) \gets (0,0,0)$\;
\For( \tcp*[f]{packets~$0$ to~$L-1$ sent}){$i \gets 0$ \textbf{to} $L-1$} {

 \If(\tcp*[f]{packet~$i$ not erased}){$e_i = 0$}{

 $\mathrm{current\_seq\_\#}\gets i$\;
      \For{$j \gets \mathrm{previous\_seq\_\#}+1$ \textbf{to} $\mathrm{current\_seq\_\#}$} {
      $\mathcal{W} \gets \{j-T, j-T+1, \ldots, j\}$\;

%

       $\bar B_j\gets \max\{\mathrm{span}(e_\mathcal{W}), \hat B_{j-1}\}$ \;
$\bar N_j\gets \max\{\mathrm{wt}(e_\mathcal{W}), \hat N_{j-1}\}$ \;
$N_{\max}\gets \max\{\mathrm{wt}(e_\mathcal{W}), N_{\max}\}$ \;

            \eIf(\tcp*[f]{trivial}){$\bar N_j=0 \mathrm{\: or\:} \bar N_j = T+1$}{
        $ (\hat B_j,\hat N_j) \gets (\hat B_{j-1},\hat N_{j-1})$\;
       }(\tcp*[f]{compute 3 hypothetic rates}){
%
       $R_B \!\gets\!\! \begin{cases} 0 &\text{if $\bar B\! =\! T+1$,}\\\mathrm{C}(T,\bar B_j,\max\{\hat N_{j-1},1\})&\text{if $\bar B\!<\!T+1$}\end{cases}$\;
$R_N \gets\mathrm{C}(T,\max\{\bar B_{j-1}, \bar N_j\},\bar N_j)$\;
$R_{\mathrm{MDS}} \gets \mathrm{C}(T,N_{\max},N_{\max})$\;
\Switch{$\max\{R_B,R_N,R_{\mathrm{MDS}}\}$}{
\uCase(\tcp*[f]{$R_B$ largest}){$R_B$}{$(\hat B_j,\hat N_j) \gets (\bar B_j, \max\{\hat N_{j-1},1\})$\;
break\;}
\uCase(\tcp*[f]{$R_N$ largest}){$R_N$}{$(\hat B_j,\hat N_j) \gets (\max\{\hat B_{j-1},\bar N_j\}, \bar N_j)$\;
break\;}
\uCase(\tcp*[f]{$R_{\mathrm{MDS}}$ largest}){$R_{\mathrm{MDS}}$}{$(\hat B_j,\hat N_j) \gets (N_{\max}, N_{\max})$\;}
}
%
%

       }

  }

 $\mathrm{previous\_seq\_\#}\gets \mathrm{current\_seq\_{\#}}$\;
}
}
\caption{Estimating conservative $B$ and $N$}\label{algorithm1}
\end{algorithm}
In addition to the explicit construction of optimal streaming codes, we also present a conservative algorithm demonstrated by Algorithm~\ref{algorithm1}. The algorithm estimates conservative coding parameters $B$ and $N$ in~$L$ channel uses. Conservative in this context implies coding parameters $B$ and $N$ that do not yield the lowest rate, where $B=N=T$ and $C(T, T, T) = \frac{1}{T+1}$. Before tracking any packet erasures, the algorithm fixes the decoding delay denoted by $T$ and the duration of the algorithm denoted by~$L$. Also, initially the channel is assumed to be ideal, i.e. introducing no erasures. Therefore, the algorithm sets the starting value of~$B$, ~$N$, which are denoted by $\hat B_{-1}$ and $\hat N_{-1}$ respectively, and~$N_{\max}$, which is the maximum number of arbitrary erasures, is set to~$0$.

Every packet transmitted at channel use~$i\in\mathbb{Z}_+$ is assumed to either reach the destination in the same channel use or be erased. In practice, packets that are dropped in the network are considered erased. Depending on the application, packets received out-of-order could either be considered erased or are reordered at the application layer. In our experiments, we do not see out-of-order packets detected, hence in our implementation we considered out-of-order packets to be erased. However, in practice if we use Real Time Protocol (RTP), then the receiver will have a reordering buffer that reorders packets received out-of-order. 

For every non-erased packet received at channel use~$i\in\mathbb{Z}_+$, the algorithm first deduces the erasure pattern $e_{\mathcal{W}}\triangleq (e_{j-T},e_{j-T+1}, \ldots, e_j)\in\{0,1\}^{T+1}$ for each sliding window~$\mathcal{W}=\{j-T, j-T+1,\ldots, j \}$ of size~$T+1$ such that $j\le i$, where an element of $e_{\mathcal{W}}$ equals~$1$ if and only if the corresponding packet is erased. Let
$
\mathrm{wt}(e_\mathcal{W}) \triangleq \sum_{\ell\in\mathcal{W}}e_\ell 
$
 and
\begin{equation*}
 \mathrm{span}(e_\mathcal{W}) \triangleq
 \begin{cases}
 0 & \text{if $\mathrm{wt}(e_\mathcal{W})=0$,} \\
 p_\text{last} - p_\text{first} + 1 & \text{otherwise,}
 \end{cases} 
\end{equation*}
 be the \emph{weight} and \emph{span} of~$e_{\mathcal{W}}$ respectively, where $p_\text{first}$ and $p_\text{last}$ denote respectively the channel use indices of the first and last non-zero elements in $e_{\mathcal{W}}$. Intuitively speaking, $\mathrm{span}(e_\mathcal{W})$ is the minimum length over all intervals that contain the support of $e_\mathcal{W}$. For each deduced erasure pattern $e_{\mathcal{W}}=(e_{j-T}, e_{j-T+1}, \ldots, e_j)$, the algorithm first calculates $\mathrm{wt}(e_{\mathcal{W}})$ and $\mathrm{span}(e_{\mathcal{W}})$, and then assign the values to $(\bar B_j, \bar N_j, N_{\max})$ according to
\begin{align*}
\bar B_j&:= \max\{\mathrm{span}(e_\mathcal{W}), \hat B_{j-1}\},\\
\bar N_j&:= \max\{\mathrm{wt}(e_\mathcal{W}), \hat N_{j-1}\},\\
\noalign{\noindent and}
N_{\max} &:= \max\{\mathrm{wt}(e_\mathcal{W}), N_{\max}\}.
\end{align*}
 Then one of the following updates will occur:
\begin{enumerate}
\item[(i)] $\bar B_j$ gets assigned to $\hat B_j$.
\item[(ii)] $\bar N_j$ gets assigned to $\hat N_j$
\item[(iii)] $N_{\max}$ gets assigned to both $\hat B_j$ and $\hat N_j$.
\end{enumerate}
To precisely explain when will each update occur, the estimates $\hat B_j$ and $\hat N_j$ will be output according to the following three mutually exclusive cases: \\
\textbf{Case} $\bar N_j = 0$:

In this case,
\[
\bar B_j=\bar N_j= \mathrm{wt}(e_{\mathcal{W}})= \mathrm{span}(e_{\mathcal{W}})= \hat N_{j-1}=\hat B_{j-1}=0,
\]
 which implies that no erasure has yet occurred upon the receipt of packet~$j$. Then, Algorithm~\ref{algorithm1} sets $\hat N_{j}=\hat B_{j}=0$, meaning that the estimates for $N$ and $B$ remain to be $0$. \\
 \textbf{Case $\bar N_j = T+1$}: 

Here all the elements of $e_{\mathcal{W}}$ equal $1$. This means that all the packets in the window $\{j-T, j-T+1, \ldots, j\}$ are erased. In this case, no $(n,k,T)_{\mathbb{F}}$-code can correct $e_{\mathcal{W}}$. Therefore, Algorithm~\ref{algorithm1} sets $\hat N_{j}=\hat N_{j-1}$ and $\hat B_{j}=\hat B_{j-1}$, i.e. the algorithm keeps the estimates of $N$ and $B$ unchanged.\\
\textbf{Case $0< \bar N_j \ne T+1$}:

In this case, every length-$(T+1)$ erasure pattern $\varepsilon^{T+1}$ that has happened up to channel use~$j$ can be categorized into the following two types, with the terminology that $\varepsilon^{T+1}$ is a~$(B,N)$-erasure sequence if either $ \mathrm{span}(e_{\mathcal{W}})\linebreak\le B$ or $ \mathrm{wt}(e_{\mathcal{W}})\le N$ holds:
\begin{itemize}
\item[(i)] $\varepsilon^{T+1}$ consists of all ones, hence it is uncorrectable;
\item[(ii)] $\varepsilon^{T+1}$ is simultaneously a $(\bar B_j, \max\{\hat N_{j-1},1\})$-erasure sequence, a $(\max\{\hat B_{j-1},\bar N_j\}, \bar N_j)$-erasure sequence, and a $(N_{\max}, N_{\max})$-erasure sequence.
        \end{itemize}
        By construction, every length-$(T+1)$ erasure pattern up to channel use $j-1$ can be either Type~(i) or is an $(\hat B_{j-1}, \hat N_{j-1})$-erasure sequence. Therefore, Algorithm~\ref{algorithm1} calculates the best estimates for $\hat B_j$ and $\hat N_j$ so that the following two conditions hold:
         \begin{itemize}
         \item[(I)] Every length-$(T+1)$ erasure pattern up to channel use~$j$ can be classified into either Type~(i) or is a $(\hat B_{j}, \hat N_{j})$-erasure sequence.
             \item[(II)] The loss in the maximum achievable rate induced by updating the estimates from $(\hat B_{j-1}, \hat N_{j-1})$ to $(\hat B_{j}, \hat N_{j})$ is minimized.
             \end{itemize}
            The presence of Condition~(II) is essential because it guarantees that the algorithm cannot output the trivial estimates $\hat B_{j} = \hat N_{j}=T$ that lead to the lowest rate~$\mathrm{C}(T,T,T)=\frac{1}{T+1}$.

            To get the best estimates of $\hat B_j$ and $\hat N_j$ so that Conditions~(I) and~(II) hold, Algorithm~\ref{algorithm1} computes three hypothetic rates based on the $(T,B,N)$-capacity as follows:
            \begin{align*}
             R_B& := \begin{cases}0 & \text{if $\bar B_j = T+1$,} \\
\mathrm{C}(T,\bar B_j,\max\{\hat N_{j-1},1\}) & \text{if $\bar B_j < T+1$,}
             \end{cases}
            \\
             R_N&  :=\mathrm{C}(T,\max\{\bar B_{j-1}, \bar N_j\},\bar N_j)\\
            \noalign{\noindent and}
            R_{\mathrm{MDS}} &:= \mathrm{C}(T,N_{\max},N_{\max})
            \end{align*}
            respectively, where $R_B$ denotes the hypothetic maximum achievable rate if $\hat B_j$ is assigned the value~$\bar B_j$ followed by $\hat N_j$ being assigned the value $\max\{\hat N_{j-1},1\}$ (note that any $(\hat B_j, \hat N_{j-1})$-erasure sequence is also a $(\hat B_j, \max\{\hat N_{j-1},1\})$-erasure sequence), $R_N$ denotes the hypothetic maximum achievable rate if $\hat N_j$ is assigned the value~$\bar N_j$ followed by $\hat B_j$ being assigned the value $\max\{\hat B_{j-1},\hat N_j\}$ (note that any $(\hat B_{j-1},\hat N_j)$-erasure sequence is also a $(\max\{\hat B_{j-1},\hat N_j\}, \hat N_j)$-erasure sequence), and $R_{\mathrm{MDS}}$ denotes the hypothetic maximum rate if both $\hat B_j$ and $\hat N_j$ are assigned the same value $N_{\max}$. Algorithm~\ref{algorithm1} sets $(\hat B_j, \hat N_j)$ as shown at the end of the pseudocode
so that the resultant maximum achievable rate~$\mathrm{C}(T, \hat B_j, \hat N_j)$ equals $\max\{R_B,R_N,R_{\mathrm{MDS}}\}$.

\smallskip
Combining the above three cases, we conclude that for all $0\le i\le L-1$, Algorithm~\ref{algorithm1} generates estimates $(\hat B_i, \hat N_i)$ such that Conditions~(I) and~(II) hold.
\subsection{Interleaved Conservative Algorithm Based on Algorithm~\ref{algorithm1}} \label{subsecInterleavedAlg}
By providing conservative estimates for $B$ and $N$, surely algorithm~\ref{algorithm1} yields a code that perfectly corrects all observed length-$(T+1)$ correctible erasure sequences. However, obviously Algorithm~\ref{algorithm1} generates a sequence of recommended coding rates that is monotonically decreasing over time. This is one noticeable concern. 

To try to solve this issue, we propose the \emph{network-adaptive algorithm} that is based on interleaving Algorithm~\ref{algorithm1} as follows: At each channel use~$\ell=0, L, 2L, \ldots$, an instance of Algorithm~\ref{algorithm1} denoted by $\mathcal{A}_\ell$ is initiated. Each $\mathcal{A}_\ell$ lasts for $2L$ channel uses, and let $(\hat B_j^{(\ell)}, \hat N_j^{(\ell)})$ denote the corresponding estimates generated at channel use~$j$. Then at each channel use~$j$, the network-adaptive algorithm outputs the estimate $(\hat B_j^{(\ell)}, \hat N_j^{(\ell)})$ provided by~$\mathcal{A}_\ell$ at channel use~$j$ where~$\ell$ is the unique integer that satisfies $\ell+L\le j< j+2\ell$.

In simple words, each interleaved Algorithm~\ref{algorithm1} will run for $2L$ channel uses where the first~$L$ estimates are ignored by the algorithm and the last~$L$ estimates are output by the network-adaptive algorithm.
Our construction guarantees that the coding rate generated by our network-adaptive algorithm is not always monotonically decreasing over time. Particularly, if there are no erasures for consecutive~$2L$ channel uses, the next estimates of $(B,N)$ would be $(0,0)$.

\section{Network-Adaptive Streaming Scheme} 
\label{scheme}
After receiving packets at the destination, the receiver can estimate the values of $\{(\hat B_j, \hat N_j)\}_{j\in\mathbb{Z}_+}$ as suggested by the network-adaptive algorithm described in Section~\ref{subsecInterleavedAlg}. The parameter estimator uses the network-adaptive algorithm to generate the estimates $(\hat B_i, \hat N_i)$ when the codeword transmitted at time $i$ is received by the destination. To minimize estimation error at the source side induced by obsolete channel information, the destination feeds back the estimates instantaneously every time it receives a packet. This way more erasure patterns can be corrected at the cost of a small rate loss.

\subsection{Code Transition}
Initially, from channel use~$0$, the network-adaptive algorithm outputs $(\hat B_0, \hat N_0)=(0,0)$. This implies that the source will use the trivial rate-one encoder, so that the codeword communicated is identical to the message produced. The source stays with the trivial rate-one encoder until the algorithm updates the estimates to positive values for $B$ and $N$.

Whenever the algorithm provides new estimates for $(B,N)$ at channel use~$j$ denoted by $(\hat B_j, \hat N_j)$, the source shifts to the new encoder associated with the code $\mathcal{C}_{T,\hat B_j, \hat N_j}$ (defined at the end of Section~\ref{construction}).
To secure a smooth transition from using an old encoder with parameters $(B_{\text{old}}, N_{\text{old}})$ to using a new encoder with parameters $(B_{\text{new}}, N_{\text{new}}) \ne (B_{\text{old}}, N_{\text{old}})$, the source has to shield every transmitted packet by protecting it by either the old or new encoder. The smooth transition is carried out as described below.

Suppose the source wants to shift to a new encoder associated with $\mathcal{C}_{T, B_{\text{new}}, N_{\text{new}}}$ starting from channel use~$i$, it will use both the old and new encoders to encode the same message into old and new codewords from channel use~$i$ to channel use~$i+T$. This will have the messages generated before channel use~$i+T$ protected by the old codewords in $\mathcal{C}_{T, B_{\text{old}}, N_{\text{old}}}$. While the messages produced from channel use~$i+T$ till the next encoder transition will be protected by the new codewords in $\mathcal{C}_{T, B_{\text{new}}, N_{\text{new}}}$. During the next encoder transition, the new encoder will be replaced by another newer encoder and be treated as an old encoder. This encoder transition procedure repeats guarding every transmitted packet. 

Since every message is protected by either an old encoder with parameters $(B_{\text{old}}, N_{\text{old}})$ or a new encoder with parameters $(B_{\text{new}}, N_{\text{new}})$ during the encoder transition, any $(B_{\text{old}}, N_{\text{old}})$-erasure sequence of length-$(T+1)$ that occurs before the transition can be corrected by the old encoder and any $(B_{\text{new}}, N_{\text{new}})$-erasure sequence of length-$(T+1)$ that occurs during and after the transition can be corrected by the new encoder.

\subsection{Prototype}\label{subsecPrototype}
The prototype of our proposed network-adaptive streaming scheme is shown in Fig.~\ref{figure1}, which is explained as follows. 

The parameter estimator at the destination uses the network-adaptive algorithm to generate the estimates $(\hat B_i, \hat N_i)$ for each~$i\in\mathbb{Z}_+$. An FEC message is generated at the source at each channel use~$i$. The FEC message consists of a data buffer, an integer specifying the size of the buffer, a sequence number and the latest available estimates $(\hat B, \hat N)$ fed back from the destination. The FEC message is then encoded into an FEC codeword and transmitted through the erasure channel. Each FEC codeword consists of a codeword buffer, an integer specifying the size of the codeword buffer, the sequence number originated from the corresponding FEC message, and the coding parameters $(\hat B, \hat N)$. 

The destination decodes all the messages generated before channel use~$i-T$ for every codeword received at channel use~$i$, that have not been decoded yet. The relevant decoder can be chosen by the destination based on the coding parameters contained in all the received codewords up to channel use~$i$. Hence, every received codeword may result in more than one decoded message. For every reconstructed FEC message, the corresponding data buffer, the size of the buffer and the sequence number are extracted for further processing at the application.

\section{Simulation and Experimental Results}
\label{experiments}
To start with, we first design an implementation of our proposed network-adaptive streaming algorithm. To experiment our network-adaptive streaming algorithm, we test our algorithm on a simulated statistical erasure channel and on real-world channels. We compare our algorithm against both uncoded, non-adaptive schemes and MDS-adaptive schemes. The performance gains are evaluated in terms of frame loss rate (FLR), Perceptual Evaluation of Speech Quality (PESQ) score, and coding rate. We report that our proposed network-adaptive streaming algorithm outperforms uncoded and non-adaptive schemes in all contexts. Finally, our network-adaptive algorithm performs either better or close compared to MDS-adaptive coding, but with always higher coding rate and hence lower redundancy.
\subsection{Implementation of Network-Adaptive Streaming Scheme} \label{sectionImplementation}
\begin{figure}[t!]
\centering
\includegraphics[width=0.85\linewidth, bb=0 0 220 140]{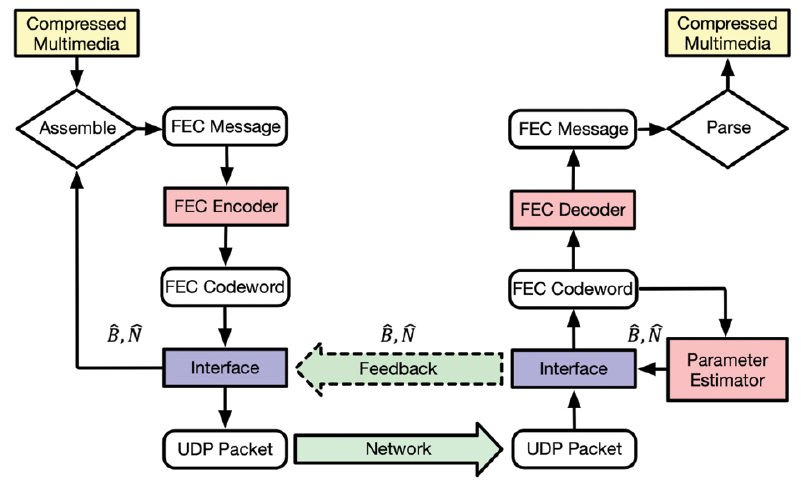}
\caption{Prototype of network-adaptive streaming scheme}
\label{figure1}
\end{figure}
To explore the potential of our proposed network-adaptive streaming scheme described in Section~\ref{scheme}, we implement the proposed scheme for low-latency communication between a source and a destination in C++ programming language.

We assume that the source transmits a stream of compressed multimedia frames over the Internet to the destination. By using a standard video codec or voice codec, each compressed multimedia frame could be created from the raw data.
Next, the compressed multimedia frame, together with the estimated coding parameters received from the feedback channel, is encapsulated in an FEC message. The FEC message is further encoded into an FEC codeword to be encapsulated in a network packet, which is then forwarded to the destination.

In all our experiments we focus on audio multimedia frames. This is because in an interactive video-streaming application such as video conferencing, voice is more delay-sensitive than video. This paper focuses on applying adaptive FEC schemes to protect the data with the most stringent delay constraint, i.e., voice. Since our application of choice in this paper is lower bandwidth audio/voice, we do not adopt congestion control in our experiments. The application of our FEC schemes on video and its interaction with congestion control schemes are interesting directions for left for future research.

Every network packet is either received by the intended destination or dropped (erased). Each FEC codeword received at the destination is extracted from every network packet received, and one or more FEC messages are recovered based on the codeword. A recovered compressed multimedia frame is extracted from every recovered FEC message, and then decompressed by the video or voice codec back to raw data.

The two interface modules between the streaming code and the network layer are illustrated in Fig.~\ref{figure1}.
The first module is at the source. The interface at the source simultaneously encapsulates every FEC codeword into a UDP packet and forwards it to the message assembler every estimated coding parameter obtained from the feedback channel. The second module is the interface at the receiver-side that concurrently extracts the codeword buffer in every network packet to form an FEC codeword and forwards each estimated coding parameters over UDP to the feedback channel.
\subsection{Parameters and Error Metrics}
We compare the uncoded, non-adaptive, MDS-adaptive coding scheme FLRs achieved with our network-adaptive streaming scheme, as described in Section~\ref{scheme}. When comparing non-adaptive and MDS-adaptive schemes with our network adaptive streaming scheme, we choose a delay constraint of $T=10$ packets. For non-adaptive schemes, we fix the coding parameters $(B,N)$. To this end, we fix the frame duration and bit rate for the compressed multimedia frame to be 10~ms and 240~kbit/s respectively, which are practical as existing audio codecs typically have frame duration 2.5 -- 60~ms and bit rate 6 -- 510~kbit/s~\cite{opus,ITU_G711}.

Consequently, every 300-byte compressed frame is generated every 10~ms. The 10~ms frame duration and the delay constraint $T$ must be carefully chosen so that the resultant playback delay $T\times 10$~ms in addition to the propagation delay must be smaller than the 150~ms delay required by ITU for interactive applications~\cite{onewayTransTime,StockhammerHannuksela2005}. For example, if the propagation delay is 100~ms, then the resultant playback delay must be less than $150~\text{ms} - 100~\text{ms} = 50$~ms, which can be achieved by choosing suitable~$T$ and frame duration such that their product is~below~50~ms.

For our experimental purpose, we assume that the propagation delay is less than 50~ms and choose $T=10$ so that the resulting playback delay $T\times 10~\text{ms} = 100$~ms besides the propagation delay is below 150~ms.
For the network-adaptive algorithm described in Section~\ref{algorithm}, we set $L=1000$. In other words, each interleaved Algorithm~\ref{algorithm1} runs for $2L\times 10\times 0.001~\text{seconds}=20$~seconds where the algorithm ignores the $L=1000$ estimates generated in the first $10$ seconds, and the next $L=1000$ estimates are produced by the algorithm.

Let $M=360$ be the number of $10$-second sessions throughout the transmission, involving a total of~$L\times M = 360000$ packets that last for one hour. In each session, $L=1000$ packets are transmitted from the source to the destination. For simplicity, let the sequence number of a packet be its channel use index, starting from~$0$ and ending at~$L \times M - 1$.

During each session $m\in\{1, 2, \ldots, M\}$, the source transmits packets with sequence number between $L(m-1)$ and $Lm-1$. For each session $m$, let $\varepsilon_m$ denote the corresponding FLR achieved by our network-adaptive streaming scheme. More precisely, $L(1-\varepsilon_m)$ is the number of FEC messages with sequence number between $L(m-1)$ and $Lm-1$ which are perfectly recovered by the destination. We will express in the next two sections our simulation and experimental results respectively in terms of the average FLR defined as
$
 \frac{1}{M}\sum_{m=1}^M \varepsilon_m
$
and the fraction of \emph{low-fidelity} sessions with FLR larger than 10\% defined as
$6
 \frac{1}{M}\sum_{m=1}^M \mathbf{1}\left\{\varepsilon_m>0.1\right\}.
$

\subsection{Simulation Results for a Three-Phase Fritchman Channel} \label{sectionSimulation}
\subsubsection{Simulated FLRs} \label{subsubsecSimulatedFLR}

\begin{figure}
\centering
\includegraphics[width = 0.8\linewidth, bb=0 0 500 250]{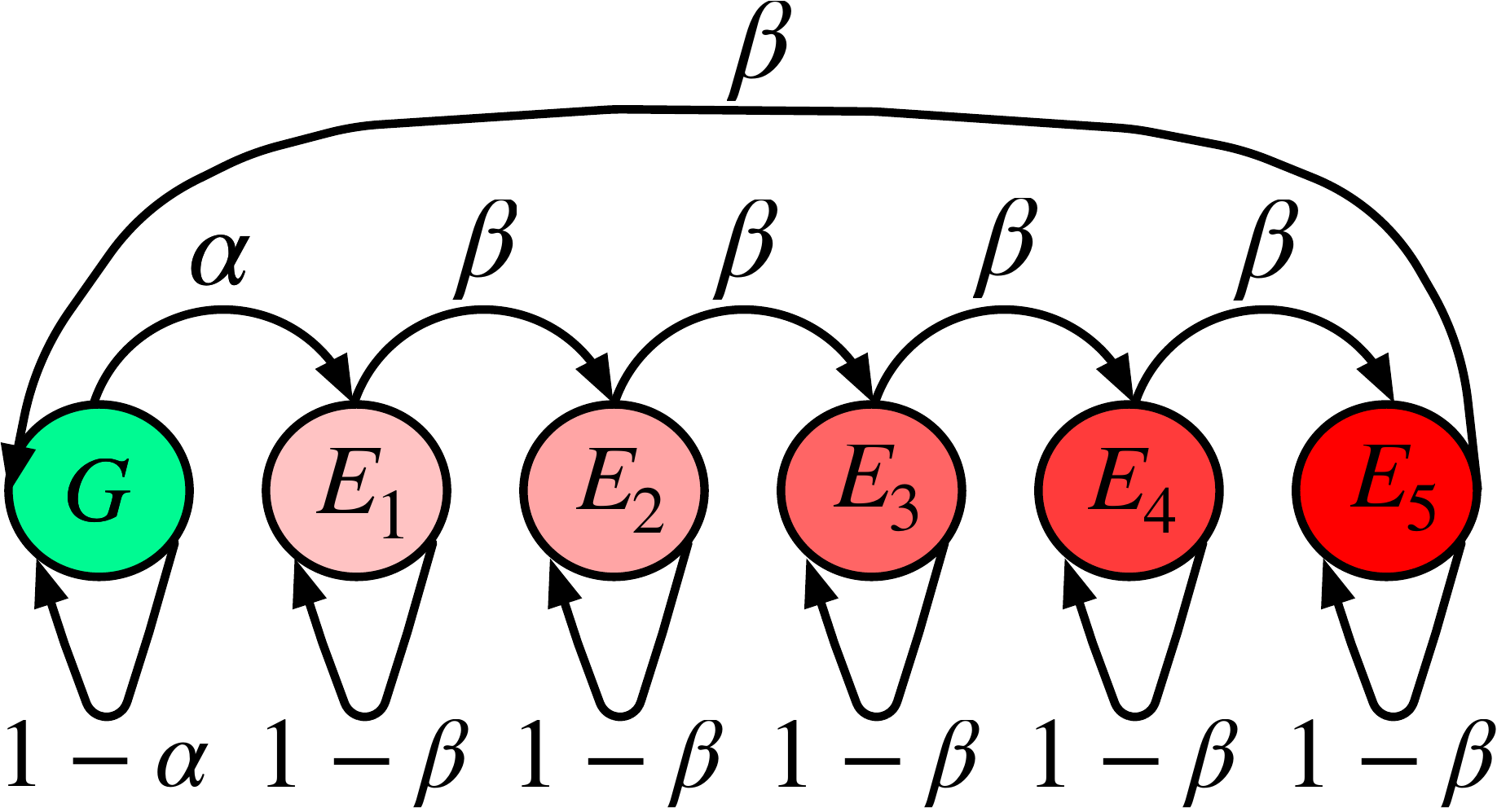}
\caption{{Six-state Markov model: Fritchman channel simulated}}
\label{fritchman-channel-model}
\end{figure}

We validate the superiority of our network-adaptive streaming scheme as described in Section~\ref{sectionImplementation} to non-adaptive streaming schemes. This is by simulating artificial packet erasures according to the Fritchman channel~\cite{fritchman-channel}, which is a well-known statistical channel that is useful for approximating packet losses experienced at the network layer~\cite{HGH2008}.

The Fritchman channel is a $(M+1)$-state Markov model which consists of one good state $G$ and $M$ bad states denoted by $E_1, E_2, ..., E_M$. In the good state $G$, each channel packet is lost with probability~$\epsilon\in[0,1)$ whereas in the bad state, each channel packet is lost with probability~$1$.

If the state at time $i$ is $G$, then the probability of transitioning to $E_1$ is $\alpha$, and the probability of remaining in the same state $G$ is $1-\alpha$ at $i+1$. When the state at $i$ is $E_M$, the transition probability to state $G$ is $\beta$, while the probability of staying in $E_M$ is $1-\beta$. If the current state is $E_l$ for some $l \in {1, 2, ..., M-1}$, then the probability of transitioning to $E_{l+1}$ is $\beta$, and it will stay in state $E_l$ with probability $1-\beta$. In our simulated Fritchman channel, we use the six-state Markov chain, where $M=5$. Fig.~\ref{fritchman-channel-model} summarizes the transition probabilities of our simulated Fritchman channel.

As long as the channel remains in bad states, the channel behaves as a burst erasure channel. The higher the number of bad states, the higher the probability of getting bursty erasures. In contrast, the channel behaves like an i.i.d.\ erasure channel when the channel stays in the good state. Hence, we expect to see more arbitrary erasures in the good state.

In our simulations, we wanted to simulate a dynamic channel, so we considered the following \textit{three-phase Fritchman channel:} The simulation of the Fritchman channel consists of three phases, where $\alpha$, $\beta$ and $\epsilon$ are fixed during the first quarter and last quarter of the simulation. While in the middle phase, the probability of being in the good state $G$ is 1. In other words, there are no consecutive bad states in the middle phase, implying that the middle phase introduces fewer burst erasures compared to the first and last phase.

For the three-phase Fritchman channel with constant parameters $(\alpha, \beta)=(0.005, 0.990)$, we focus on the case where $T=10$ for all the graphs we show in this section. For the case where $\epsilon=0.001, 0.001$, the empirical distribution of burst lengths is shown in Fig.~\ref{histogram_FE}. We exhibit a bimodal distribution with clusters at burst lengths 1 and 5.

We plot in Fig.~\ref{mean-FLR-FE} the average FLRs against the varying parameter $\epsilon$ for the uncoded scheme, the network-adaptive streaming scheme, and the best non-adaptive (fixed-rate) streaming code~$\mathcal{C}_{T, B, N}$. The coding rate of the best non-adaptive (fixed-rate) streaming code does not exceed the average coding rate of the adaptive scheme. We achieve this by getting the corresponding value/s of $(B, N)$ to the average coding rate of our network adaptive streaming code. Several values of $(B,N)$ can correspond to the same coding rate. To choose the best non-adaptive streaming code, we try running all the corresponding fixed values of $(B,N)$ on the same packet loss trace file and choose the $(B, N)$ parameters resulting in the least averaged FLR over all the sessions.

The correspondent coding rates for each $\epsilon$ are plotted in Fig.~\ref{coding-rate-FEc}, where the coding rate for the no-coding scheme is always one and thus not plotted. Fig.~\ref{mean-FLR-FE} and~\ref{coding-rate-FEc} show that compared to non-adaptive (fixed-rate) schemes, our adaptive scheme achieves approximately 10$\times$ lower FLRs and slightly higher coding rates across all values of $\epsilon$ between 0.0001 and 0.001. The gain of the adaptive scheme compared to fixed-rate codes is attributed to the significantly improved estimation of instantaneous channel conditions, and hence instantaneous change in coding parameters.

\begin{figure}[t]
\centering
\includegraphics[width=\linewidth]{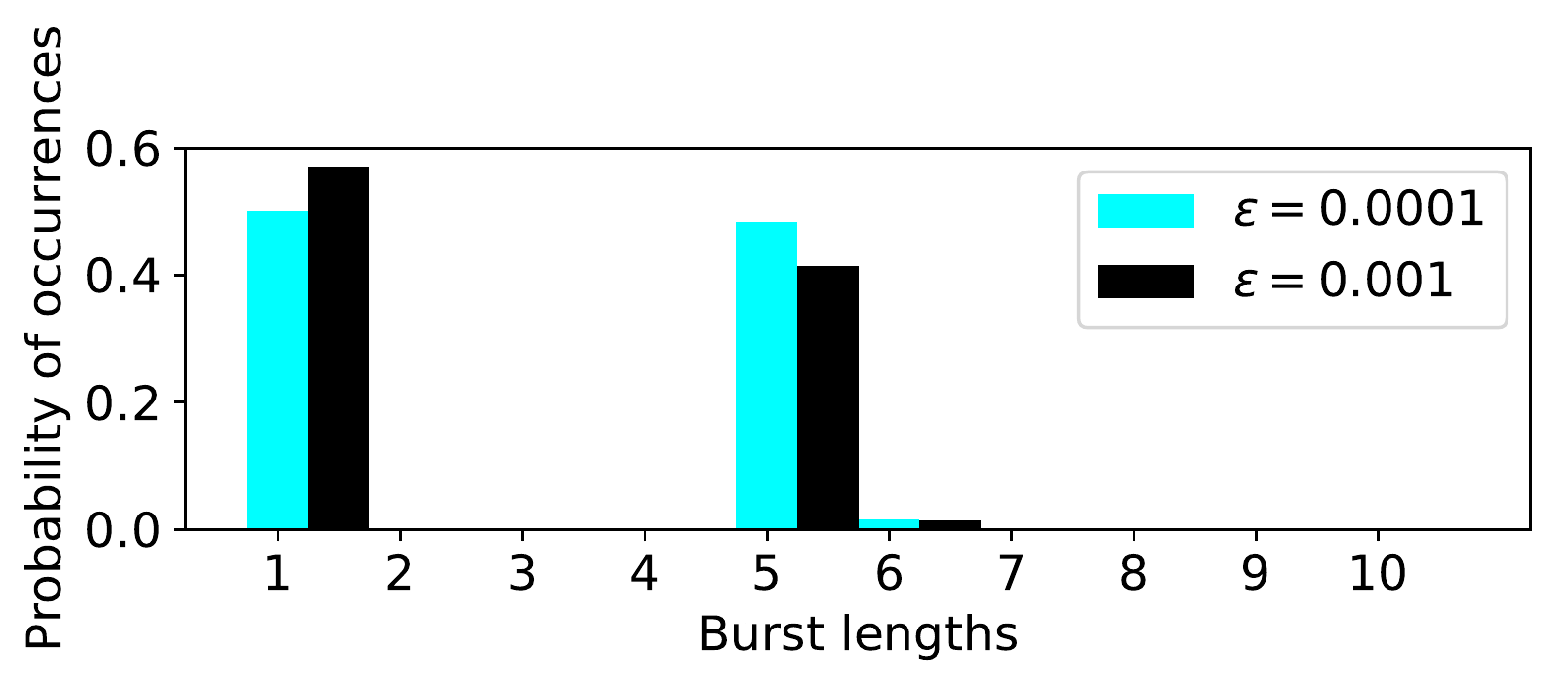}
\caption{Empirical burst-length distribution for $\epsilon=0.0001$ and $0.001$} \label{histogram_FE}
\end{figure}

\begin{figure}
\centering
\includegraphics[width=0.85\linewidth]{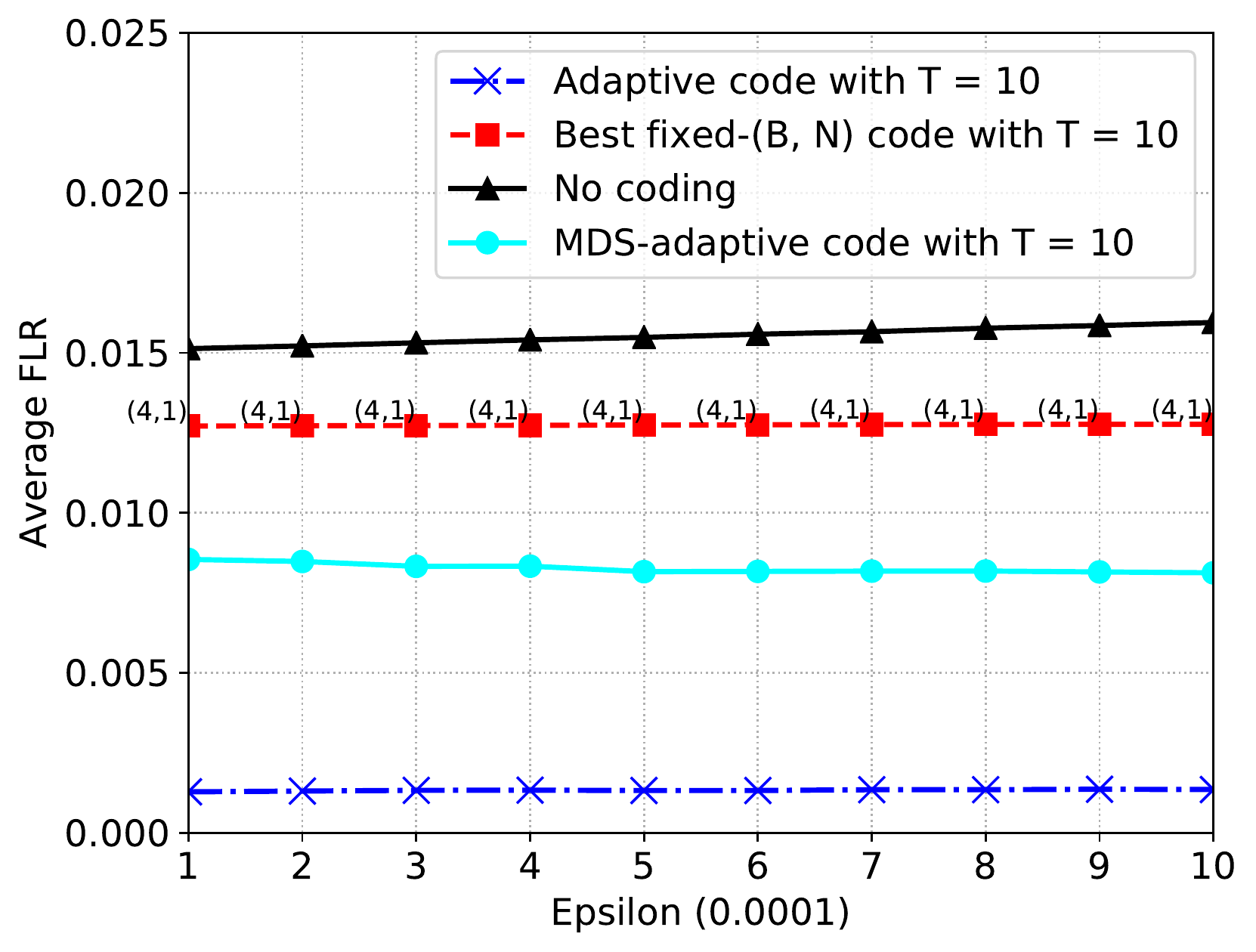}
\caption{{Average FLR obtained from Fritchman channel}}
\label{mean-FLR-FE}
\end{figure}

\begin{figure}[t]
\centering
\includegraphics[width=0.85\linewidth]{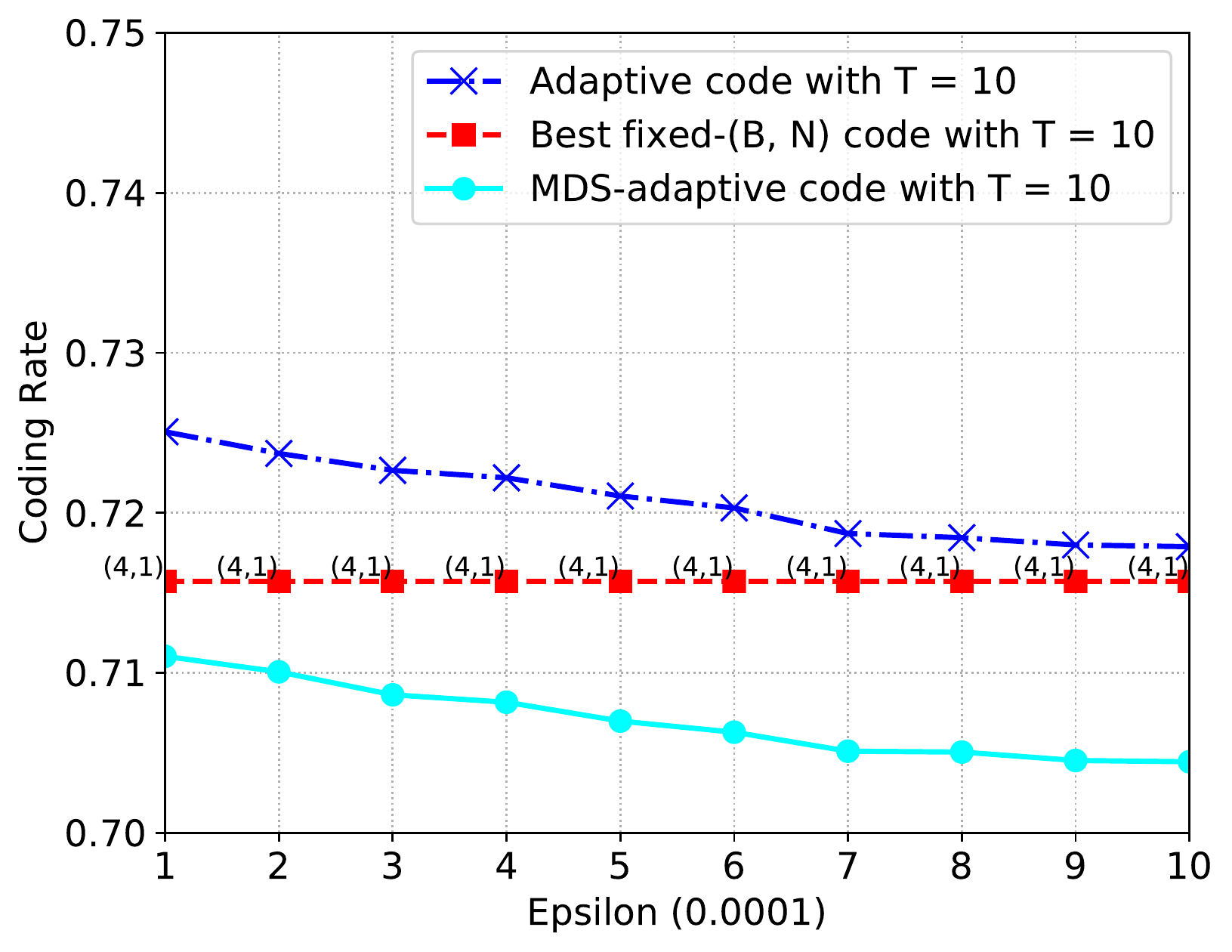}
\caption{Average coding rate} \label{coding-rate-FEc}
\end{figure}

\begin{figure}[t!]
\centering
\begin{minipage}[b]{\linewidth}
\centering
\includegraphics[width =0.85\linewidth]{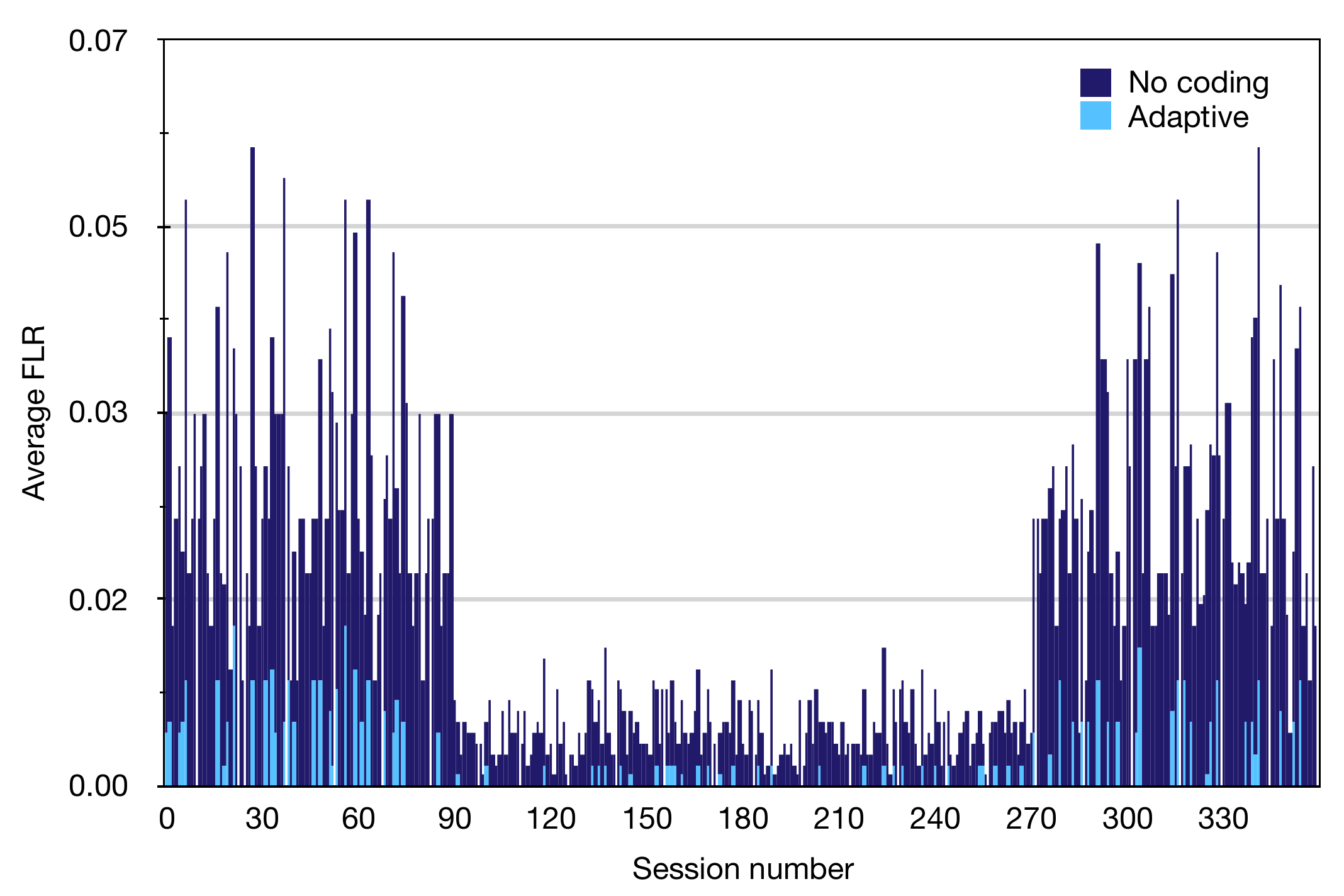}
\end{minipage}%
\caption{{Average FLRs for adaptive FEC over time for $\epsilon=0.0001$}}  \label{FLR-sessions-FE}
\end{figure}
\begin{figure}[t!]
\centering
\includegraphics[width =0.85\linewidth]{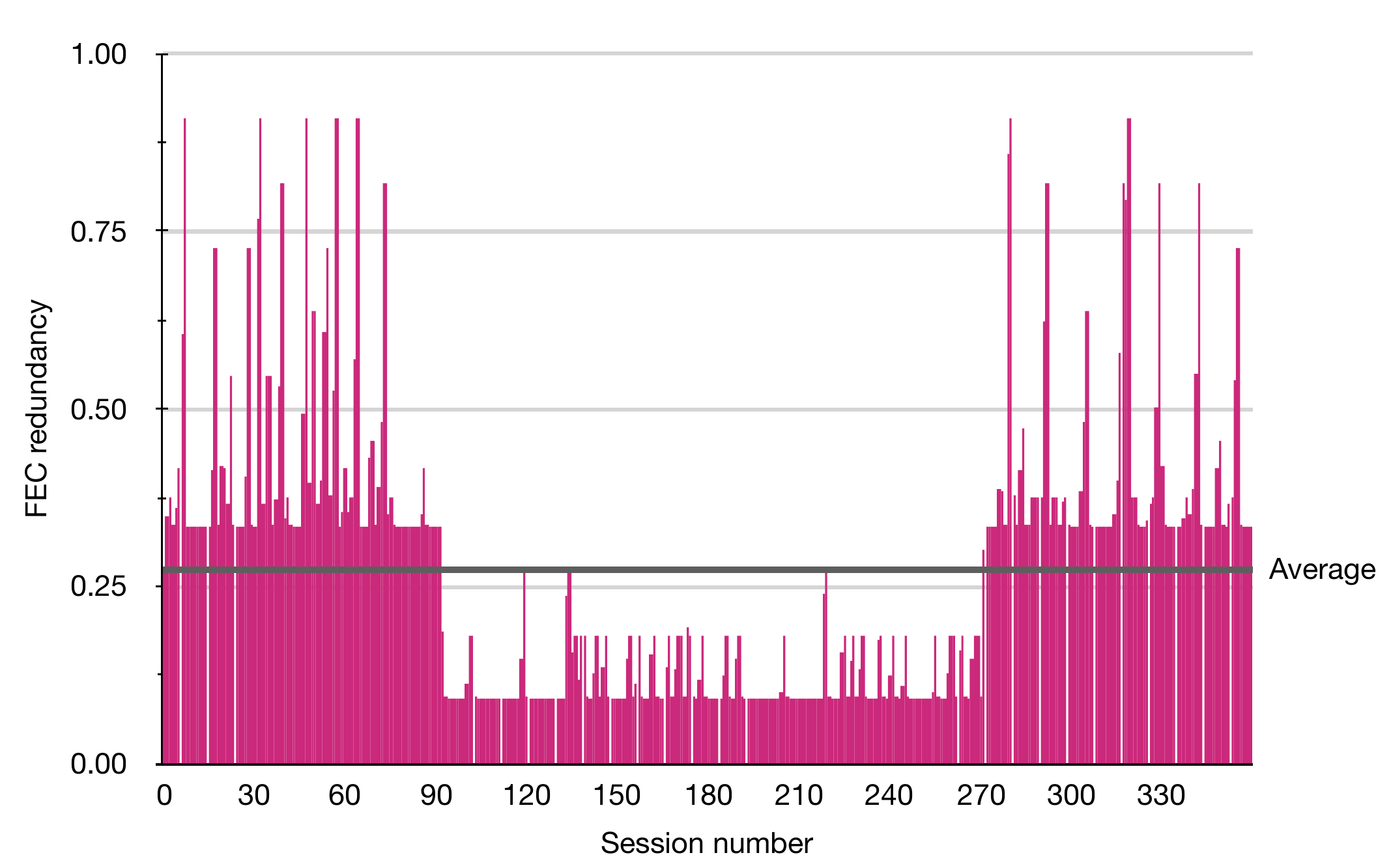}
\caption{{FEC redundancy over time for $\epsilon=0.0001$}}\label{FEC-redundancy-FE-sessions}
\end{figure}

\begin{figure}
\centering
\includegraphics[width =\linewidth]{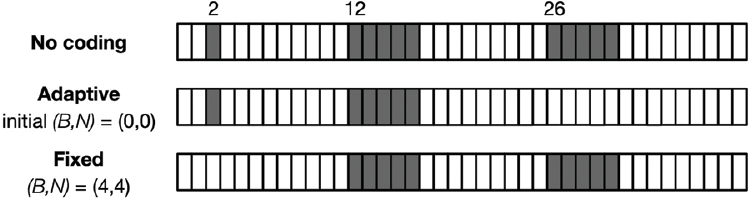}
\caption{Packet losses recovered by different schemes}
\label{figureCodeAdaptation}
\end{figure}

We also show in Fig.~\ref{FLR-sessions-FE} the variation of average FLRs for our adaptive streaming scheme and uncoded scheme across the $360$ sessions, each lasting for 10 seconds, for $\epsilon=0.0001$. It can be seen from Fig.~\ref{FLR-sessions-FE} that our adaptive scheme achieves less than half of the no-coding scheme loss rate for all sessions. In addition, we display the variation of the FEC redundancy (i.e., one minus coding rate) for our adaptive scheme in Fig.~\ref{FEC-redundancy-FE-sessions}, which demonstrates how quickly it reacts to erasures, especially at the transition between the first and middle phases and the transition between the middle and last phases.

Our simulation results reveal that our adaptive code adapts significantly better than the other state-of-the-art strategies when channel dynamics occur. Since a feedback packet is sent instantaneously when every packet is successfully received as described in earlier, the estimation error of the coding parameters due to channel dynamics is minimized.

The reason why our adaptive scheme significantly outperforms non-adaptive ones can be explained with the help of Fig.~\ref{figureCodeAdaptation}. Suppose 11 out of 40 of the network packets are dropped as shown in Fig.~\ref{figureCodeAdaptation}. Our adaptive coding scheme updates the code in this order: $\mathcal{C}_{10,1,1}$ and $\mathcal{C}_{10,5,2}$ before transmitting packets~4 and 18 respectively. Therefore, the subsequent five packets losses are all recovered by $\mathcal{C}_{10,5,2}$ as shown in Fig.~\ref{figureCodeAdaptation}, whereas the fixed-rate code $\mathcal{C}_{10,4,4}$ can only recover one packet.

\subsubsection{Simulated PESQ Scores} \label{subsubsecPESQ}
\begin{figure*}[t!]
\centering
\includegraphics[width=\linewidth]{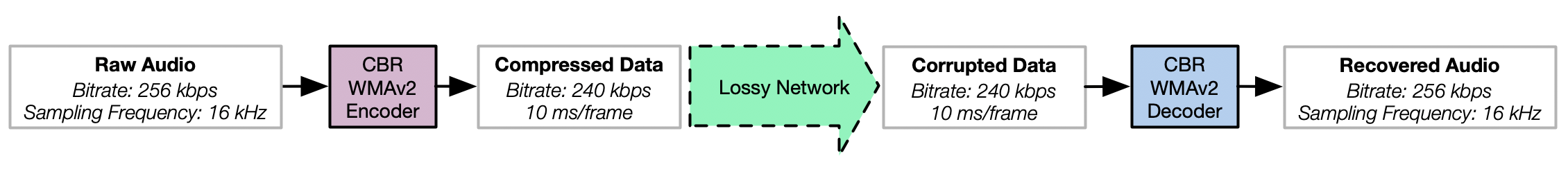}
\caption{Codec operations}
\label{figureCodec}
\end{figure*}

In the previous subsection, we presented our simulation results in terms of FLRs for compressed multimedia frames where the duration and bit rate for each compressed frame are 10~ms and 240~kbit/s respectively. {In this subsection, we use the commonly adopted wideband PESQ score~\cite{ITU_G862.2} for uncompressed multimedia (WAV) files to demonstrate the advantage in perceptual  quality of using our network-adaptive streaming scheme over uncoded and non-adaptive streaming codes.}

We first choose a one-hour uncompressed speech file with sampling frequency of 16000~Hz and sample size of 16~bits and use the constant-bitrate (CBR) WMAv2 codec to generate compressed multimedia data with sampling frequency of 16000~Hz and bit rate of~240~kbit/s. This is considerably high bit-rate to ensure any degradations comes from loss rather than compression. Then, we use the network-adaptive, uncoded and best non-adaptive scheme to transmit the compressed multimedia through the three-phase Fritchman channel as described in the previous subsection. Whenever a 300-byte compressed multimedia frame cannot be recovered by the destination, the lost frame is replaced with an 300-byte all-zero frame. The schematic diagram for the codec operations is shown in Fig.~\ref{figureCodec}.

\begin{figure}[t!]
\centering
\includegraphics[width=\linewidth]{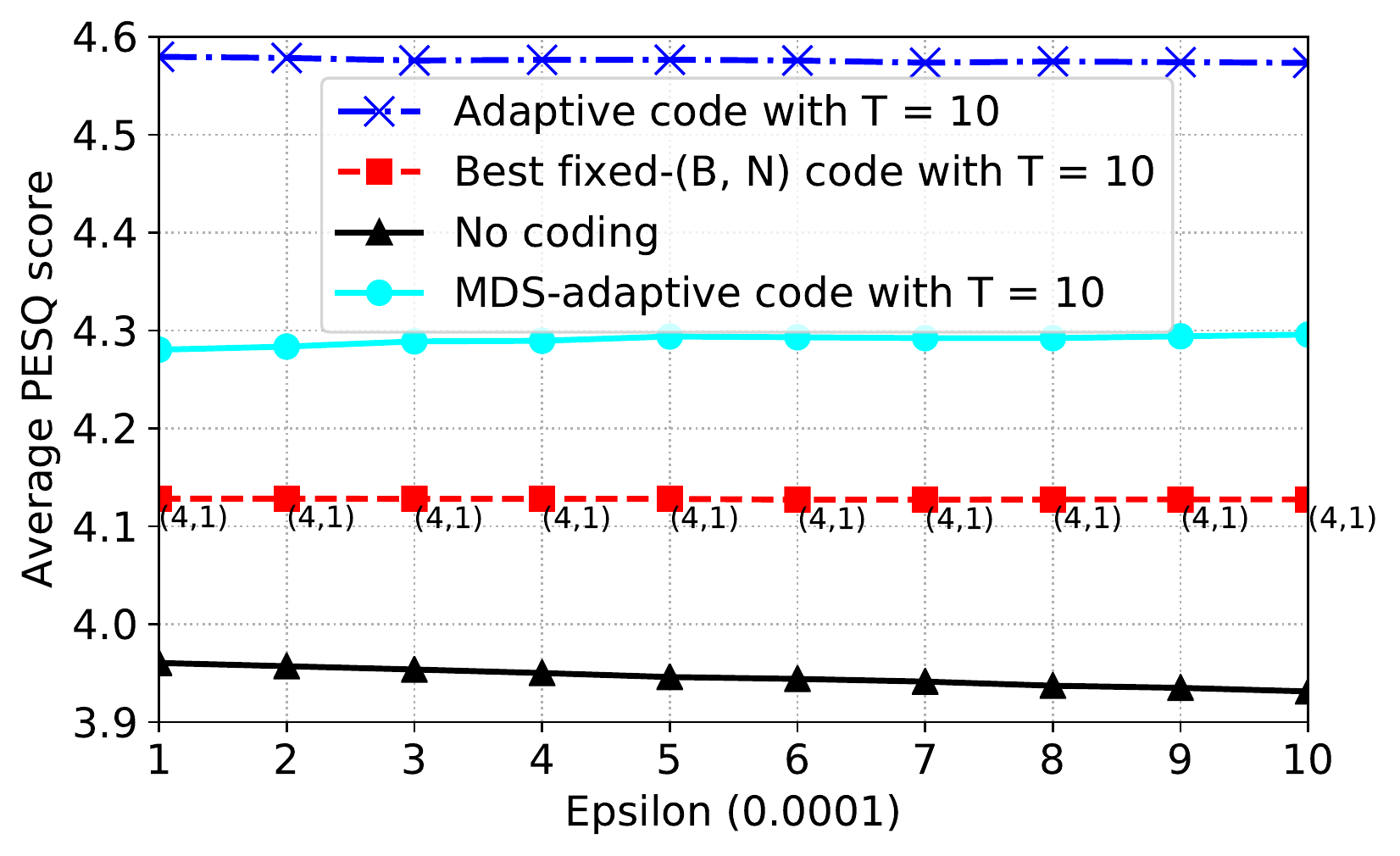}
\caption{{Average PESQ score obtained from Fritchman channel simulation}}
\label{figureLossRatePESQ_FE}
\end{figure}

\begin{figure}[t!]
\centering
\includegraphics[width=\linewidth]{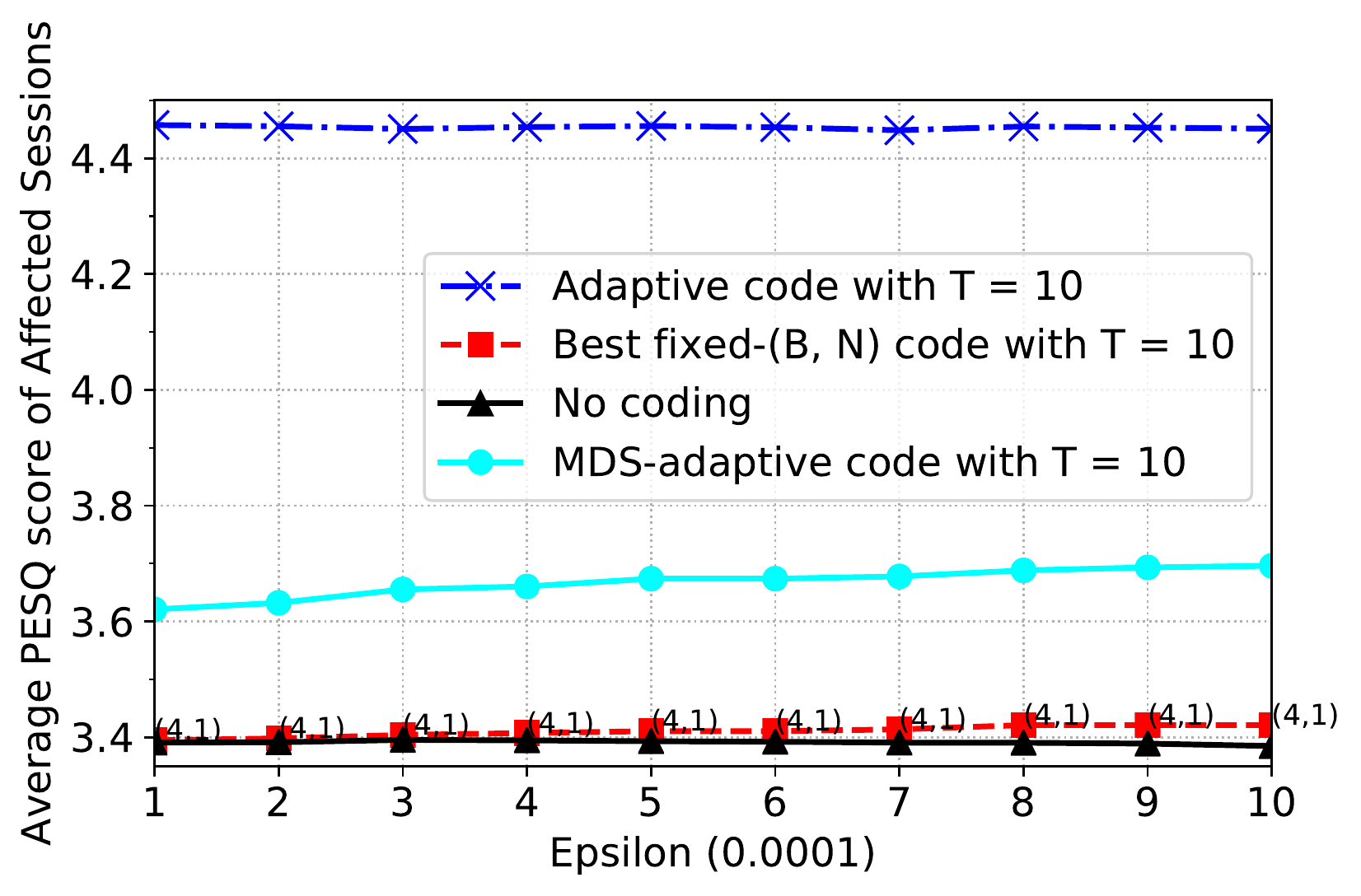}
\caption{{Average PESQ score of affected sessions obtained from Fritchman channel simulation}}
\label{figureLossRateAffectedPESQ_FE}
\end{figure}

\begin{figure}[t!]
\centering
\includegraphics[width=\linewidth]{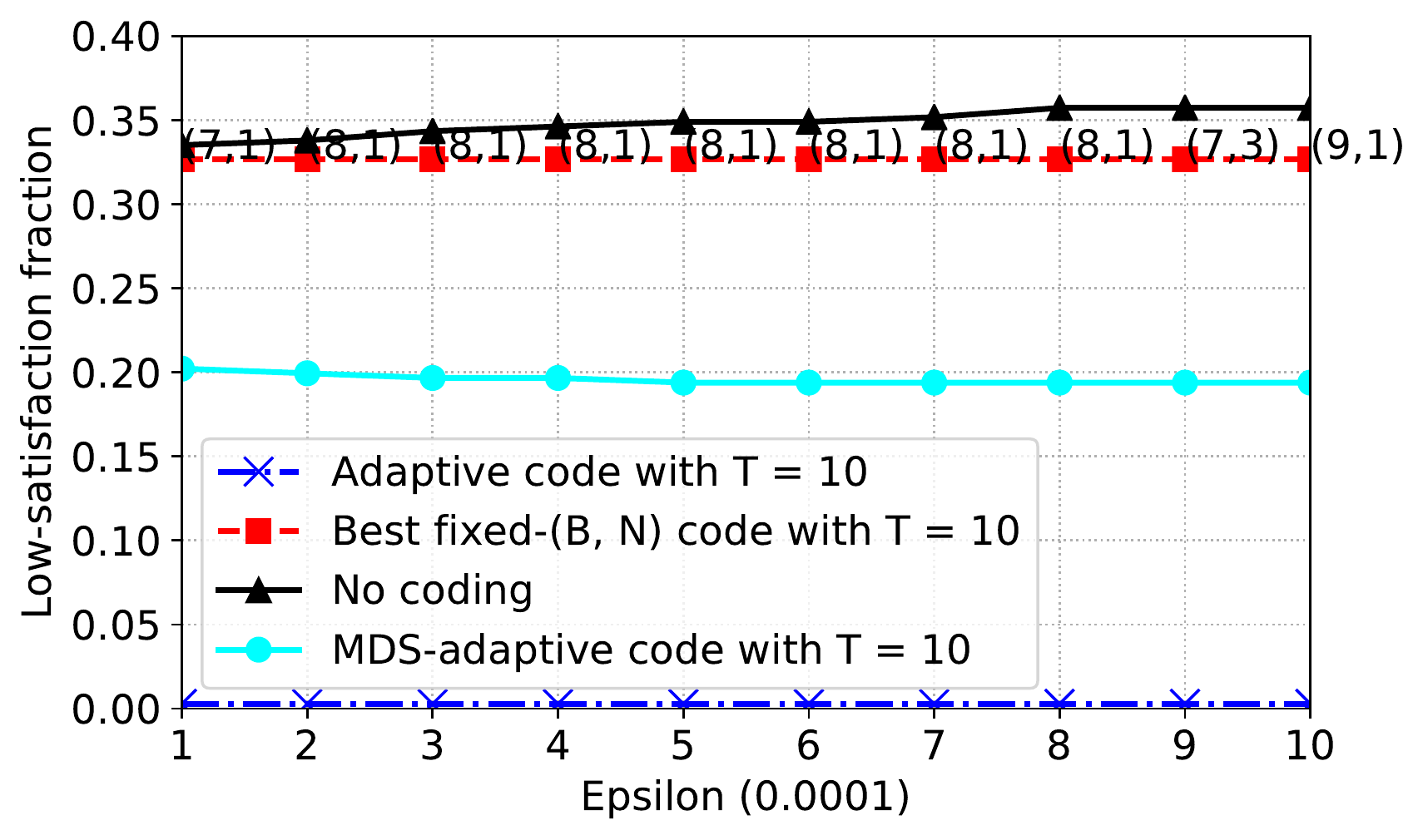}
\caption{{Low-satisfaction fraction obtained from Fritchman channel simulation}}
\label{figureLowFidelityPESQ_FE}
\end{figure}

For each 10-second session, a PESQ score is computed between the original uncompressed WAV audio and the recovered WAV audio for all the schemes. {Each PESQ score ranges from 1 to 5 where a higher score means a better speech quality~\cite{ITU_G862.2}.} We plot in Fig.~\ref{figureLossRatePESQ_FE} the average simulated PESQ scores over $M = 360$ sessions for our network-adaptive streaming scheme, the uncoded scheme and the best non-adaptive streaming code~$\mathcal{C}_{T, B, N}$ whose coding rate does not exceed the average coding rate of the network-adaptive scheme.

We can see from Fig.~\ref{figureLossRatePESQ_FE} that our adaptive streaming scheme achieves a significantly higher average PESQ score than uncoded and non-adaptive streaming schemes. However, the average PESQ score over 360 sessions may not be an indicator for the performance of our network-adaptive streaming code in affected sessions, where it suffers from higher loss rates. Hence, in Fig.~\ref{figureLossRateAffectedPESQ_FE}, we plot the average PESQ scores of only affected sessions in uncoded scheme by varying $\epsilon$ comparing non-adaptive scheme, uncoded scheme with our network adaptive scheme. Affected sessions are where the PESQ score is below 3.8 if we were using uncoded schemes. In other words, we are comparing different coding schemes only in catastrophic situations.

In Fig.~\ref{figureLossRateAffectedPESQ_FE}, we can see that in catastrophic situations, probably where the burst length is high, the performance of non-adaptive and uncoded coding scheme is almost the same. This is due to a small value of $B$, when the burst distribution is mainly at 1 and 5 as shown in Fig.~\ref{histogram_FE}. However, our network adaptive scheme is better than uncoded and non-adaptive coding schemes.

\begin{figure}[t!]
\centering
\includegraphics[width = .85\linewidth]{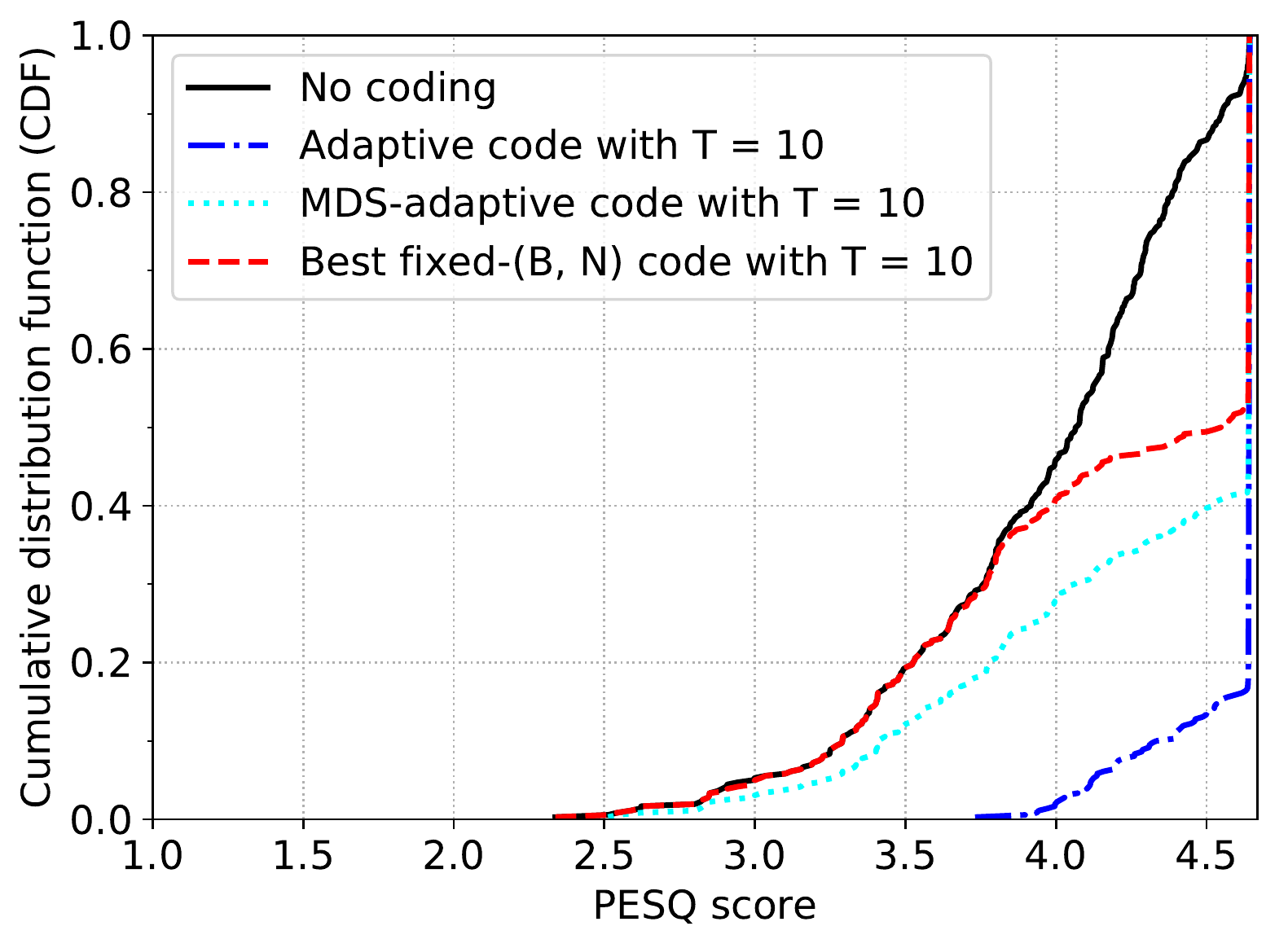}
\caption{{The CDF of simulated PESQ score for adaptive FEC, MDS-adaptive FEC and non-adaptive FEC for $\epsilon=0.0001$}}  \label{CDF_FE_0001}
\end{figure}

We plot in Fig.~\ref{CDF_FE_0001} the cumulative distribution function (CDF) of PESQ score (i.e., the fraction of the 360 sessions whose scores are less than a certain PESQ score) for each of the following schemes when $\epsilon = 0.0001$: Our adaptive streaming scheme, the best non-adaptive streaming code and the uncoded scheme. It can be seen from Fig.~\ref{CDF_FE_0001} that our adaptive scheme provides the best user experience compared to the uncoded and non-adaptive schemes. If the PESQ score of a session is lower than 3.8, many users will be dissatisfied by the unclear speech, and we will call such a session a \emph{low-satisfaction} session. Fig.~\ref{figureLowFidelityPESQ_FE} shows that our adaptive streaming scheme significantly reduces the fraction of low-satisfaction sessions compared to the uncoded and non-adaptive schemes.

\subsubsection{Optimal Streaming Codes vs.\ MDS-Based Streaming Codes} \label{subsubsectionMDSstreaming}
In order to demonstrate the advantage of using our constructed streaming codes described in Section~\ref{construction} over traditional MDS-based codes, we consider the following \emph{MDS-adaptive streaming scheme}: Instead of outputting the coding parameters $(\hat B, \hat N)$ for an optimal block code which corrects a length-$\hat B$ burst erasure and~$\hat N$ arbitrary erasures, the adaptive algorithm outputs a single coding parameter~$N$ of an MDS code which corrects only~$N$ arbitrary erasures such that the resultant coding rate~$\mathrm{C}(T, N, N)=\frac{T-N+1}{T+1}$ satisfies
\begin{align*}
\mathrm{C}(T, N, N) \le \mathrm{C}(T, \hat B, \hat N) \le \mathrm{C}(T, N-1, N-1),
\end{align*}
meaning that the resultant coding rate is the largest possible rate achieved by an MDS code that is less than $\mathrm{C}(T, \hat B, \hat N) \triangleq
\frac{T-\hat N+1}{T-\hat N+\hat B+1}$.

We plot in Fig.~\ref{mean-FLR-FE} the FLRs achieved by the MDS-adaptive streaming scheme against $\epsilon$ for the MDS-adaptive streaming scheme, and plot in Fig.~\ref{coding-rate-FEc} their corresponding coding rates. Fig.~\ref{mean-FLR-FE} and~\ref{coding-rate-FEc} show that compared to the MDS-adaptive scheme, our adaptive scheme achieves approximately 8$\times$ lower FLR and higher coding rate across all values of $\epsilon$ between 0.0001 and 0.001.

Fig.~\ref{figureLossRatePESQ_FE} shows that our optimal adaptive scheme achieves a higher PESQ score over MDS-adaptive schemes. As mentioned earlier, the average PESQ score for affected sessions gives a better indication of the performance of our scheme in worse environments, and our adaptive scheme has a 20\% improvement in PESQ score compared to adaptive MDS-streaming code in affected sessions as shown in Fig.~\ref{figureLossRateAffectedPESQ_FE}. Also, the number of affected sessions is far lower using our adaptive scheme compared to MDS-code as per Fig.~\ref{figureLowFidelityPESQ_FE}. This is also illustrated in the CDF plot in Fig.~\ref{CDF_FE_0001} that shows our adaptive network scheme having higher fraction of sessions with higher PESQ scores.

It is possible that our-adaptive network scheme chooses an MDS code if it is optimum. The superiority of our adaptive scheme to the MDS-adaptive scheme is because the fraction of non-MDS codes chosen by our adaptive algorithm is approximately 40\% or more for all $\epsilon\in[0.0001, 0.001]$. This implies that the optimal streaming codes chosen by our adaptive scheme are non-MDS more than 40\% of the time. Consequently, the streaming codes chosen by the MDS-adaptive streaming scheme are inferior to those chosen by our adaptive scheme.

On another level, the MDS scheme can definitely achieve the same level of protection as our network adaptive scheme, but at the cost of higher decoding delay $d_{\mathrm{d}}$. For example, for $\epsilon = 0.0001 - 0.001$, our network adaptive scheme can achieve similar FLRs (within 10\%) if $T$ was set to 4. Therefore, our network adaptive scheme can recover packets with a lower delay compared to MDS-based schemes.

\subsection{Experimental Results for a Wi-Fi Network}\label{sectionExperiment}
\subsubsection{Experimental FLRs} \label{subsubFLR_experiment}
\begin{figure}[t!]
\centering
\begin{minipage}[b]{\linewidth}
\centering \includegraphics[width = 0.8\linewidth]{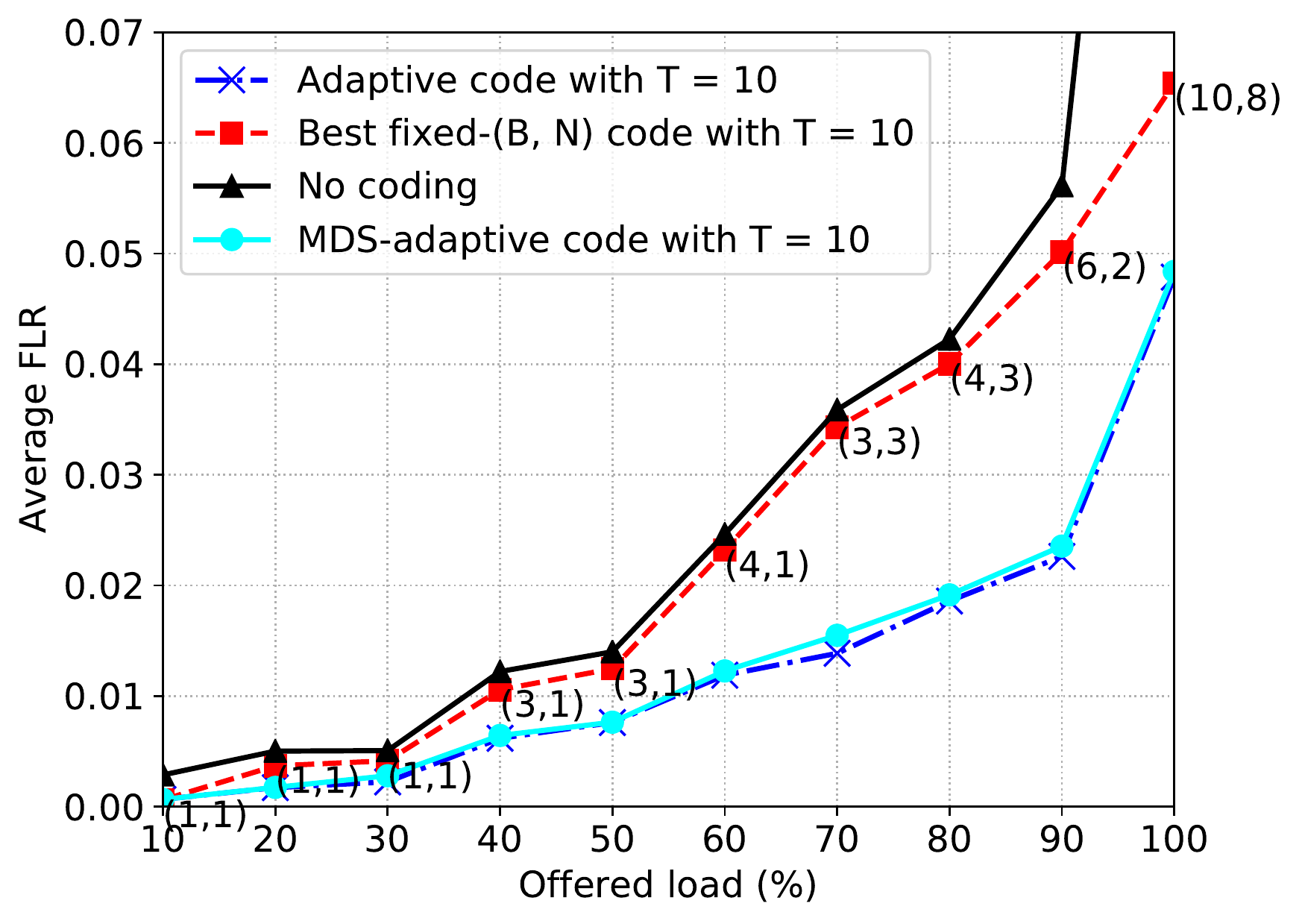}
\subcaption{Average FLR}
\label{figureLossRate}
\end{minipage}%
\\
\begin{minipage}[b]{\linewidth}
\centering \includegraphics[width =0.8 \linewidth]{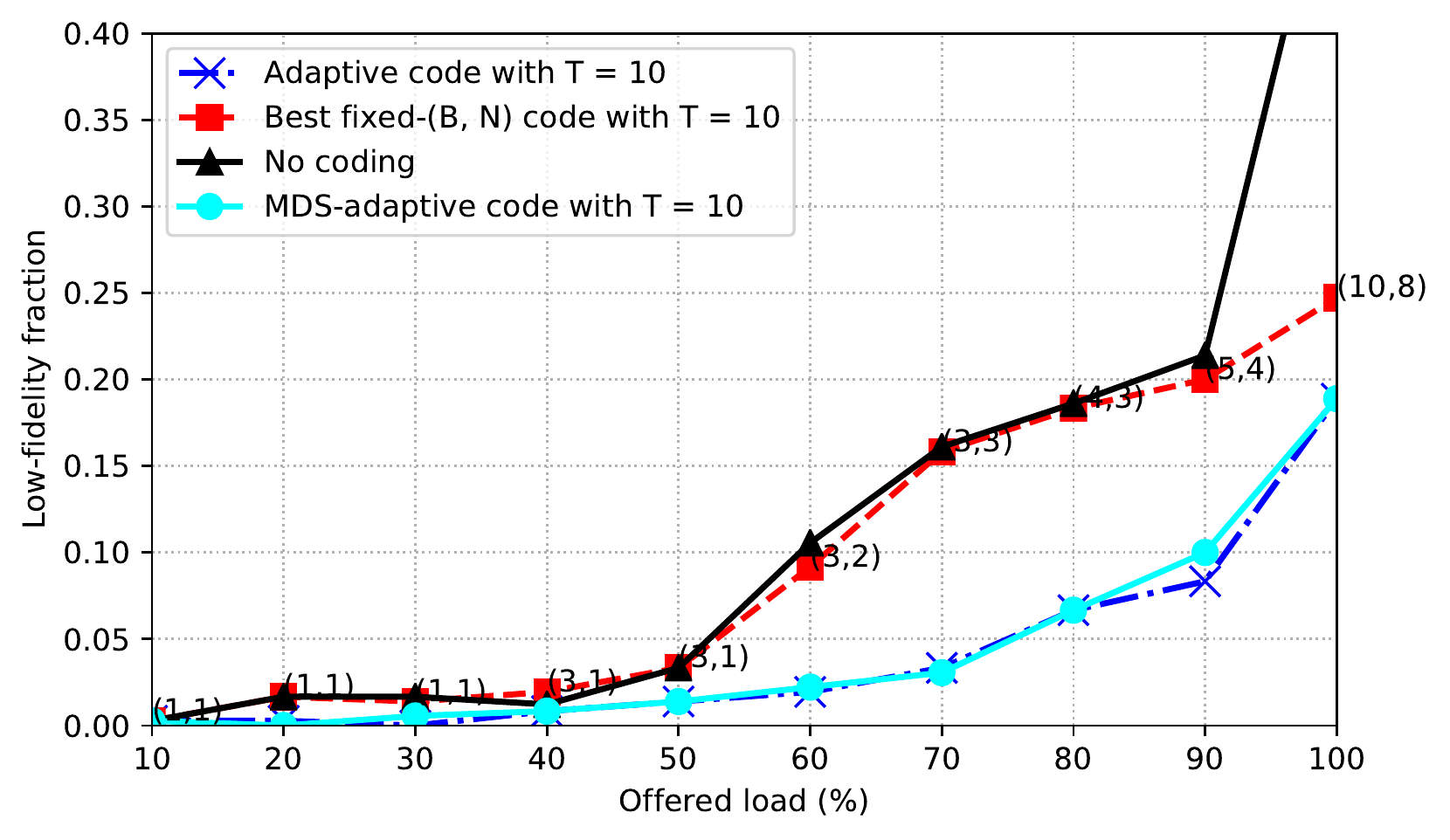}
\subcaption{Low-fidelity fraction}
\label{figureLowFidelity}
\end{minipage}
\\
\begin{minipage}[b]{\linewidth}
\centering
\includegraphics[width = 0.8\linewidth]{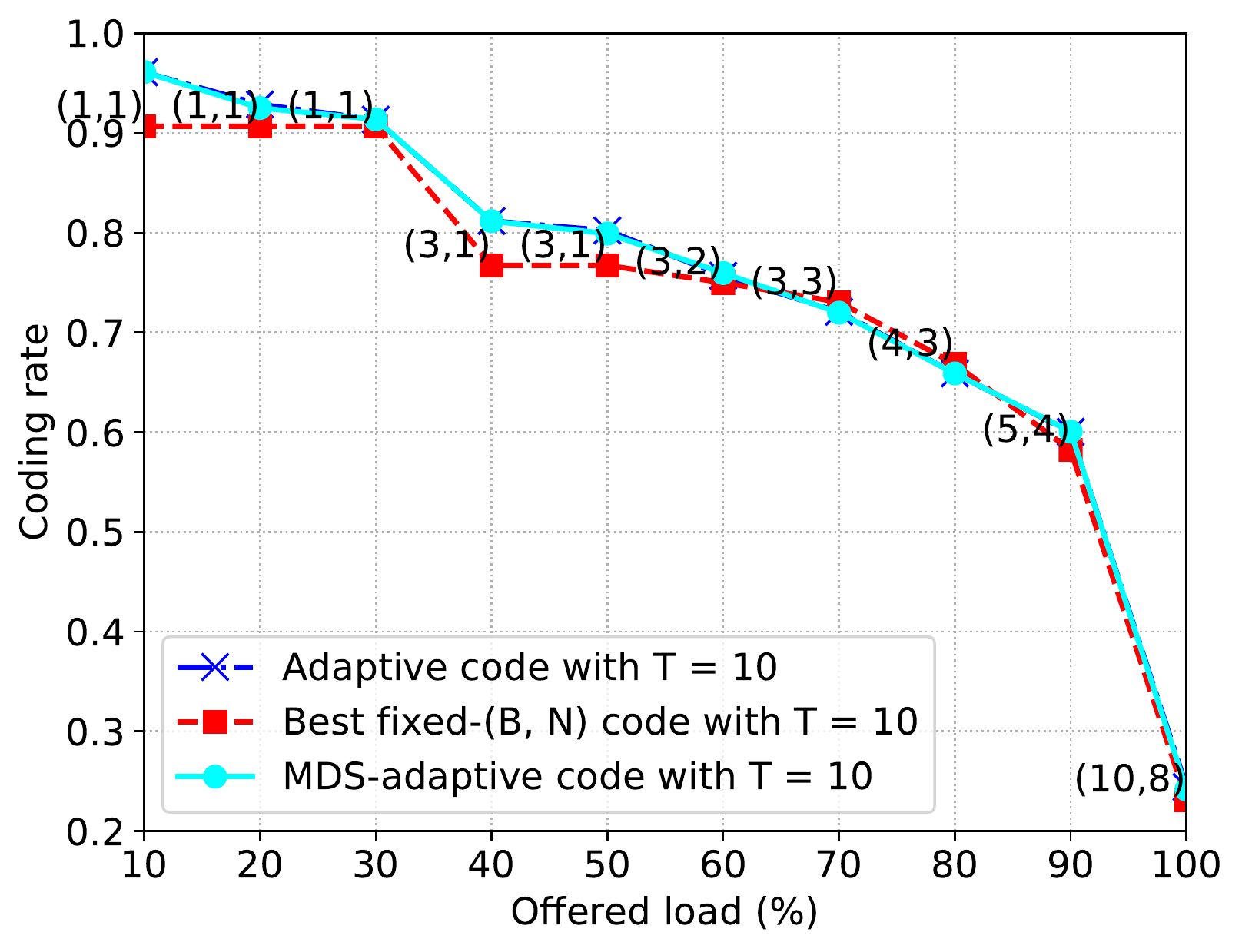}
\subcaption{Average coding rate}
\label{figureCodingRateWiFi}
\end{minipage}%
\caption{Experimental results for different streaming schemes interfered by Iperf cross traffic}
\end{figure}
In our real-world experiment, the source and the destination are connected over the same Wi-Fi network whose capacity is approximately 30~Mbit/s. To simulate cross traffic any user would experience while sharing the same Wi-Fi network, we introduce UDP cross traffic using Iperf. We call the UDP cross traffic \textit{offered load}, whose throughput is fixed at the beginning of the experiment and kept unchanged during the experiment.
The average FLRs for our network-adaptive streaming scheme as described in Section~\ref{sectionImplementation} and the uncoded scheme are plotted against the ratio of the offered load occupied by Iperf traffic in Fig.~\ref{figureLossRate}.

{To simulate the packet losses experienced during the real-world experiments, we record packet loss traces. These traces are used to compare the performance of our adaptive streaming with the best non-adaptive (fixed rate) streaming code~$\mathcal{C}_{T, B, N}$ whose coding rate does not exceed the average coding rate of the adaptive scheme.} The average FLR for the best non-adaptive streaming code~$\mathcal{C}_{T, B, N}$ with parameters $(B,N)$ is plotted in Fig.~\ref{figureLossRate}. Fig.~\ref{figureLossRate} and Fig.~\ref{figureCodingRateWiFi} show that our adaptive streaming scheme achieves significantly lower average FLRs and higher average coding rate than non-adaptive streaming codes in all offered loads. The gain of the adaptive scheme compared to fixed-rate codes is attributed to the significantly improved estimation of instantaneous channel conditions. In real-world networks, errors occur in bursty manner, meaning that most of the time the networks are free of error. Fixed-rate codes compared with adaptive codes use a larger overhead to transmit the same amount of data because it cannot adapt to a higher coding rate when the channel is error-free.

For interactive audio, low-fidelity sessions lead to unclear speech or even call termination which directly affects user experience. Fig.~\ref{figureLowFidelity} shows that our adaptive streaming scheme provides a substantially better audio quality than the uncoded and non-adaptive streaming scheme.

\begin{figure}
\centering
\includegraphics[width = 72mm]{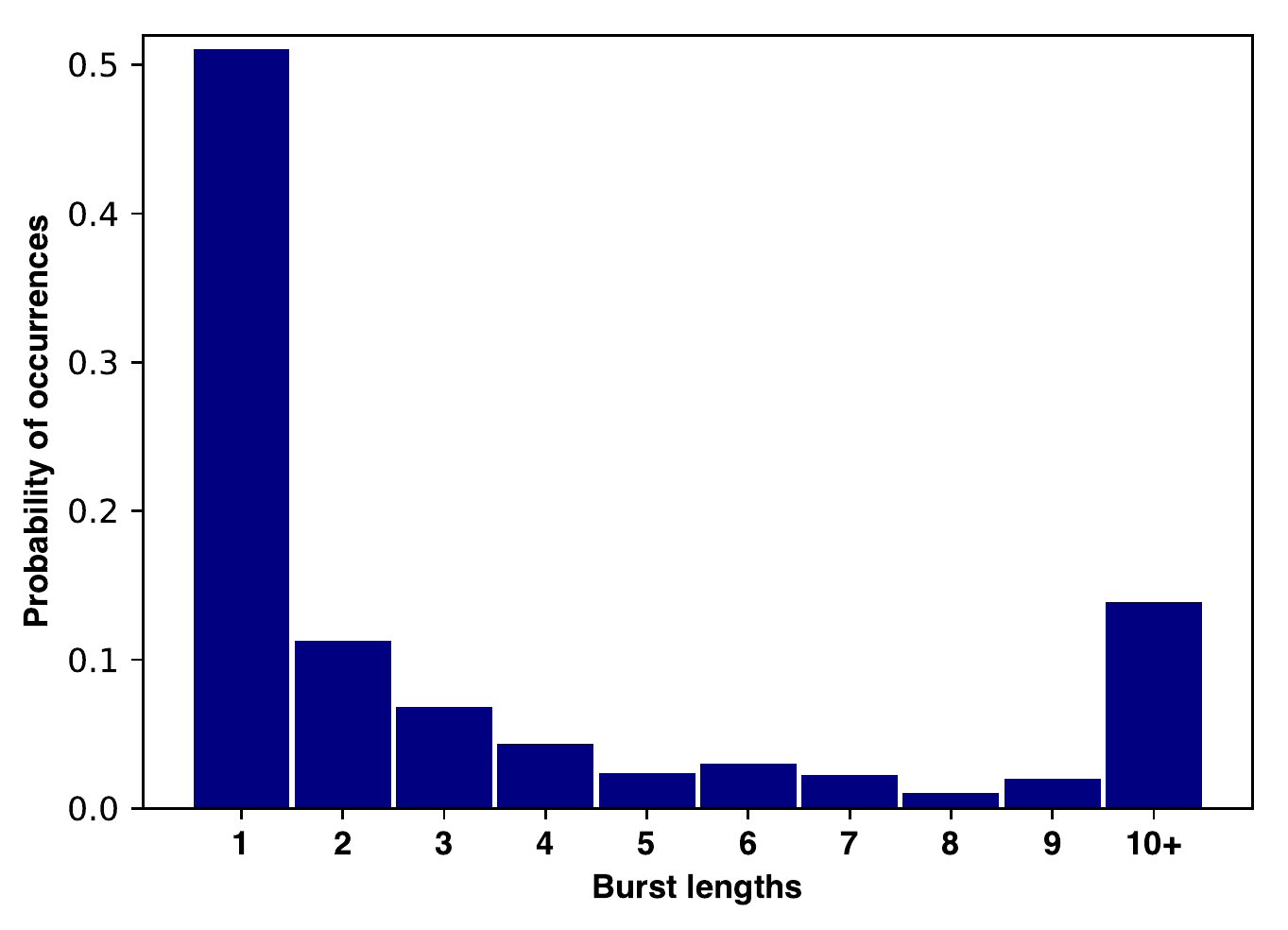}
\caption{{Empirical burst-length distribution for 40\%-capacity offered load}}
\label{histogram_wifi}
\end{figure}

\begin{figure}[t!]
\centering
\includegraphics[width = \linewidth]{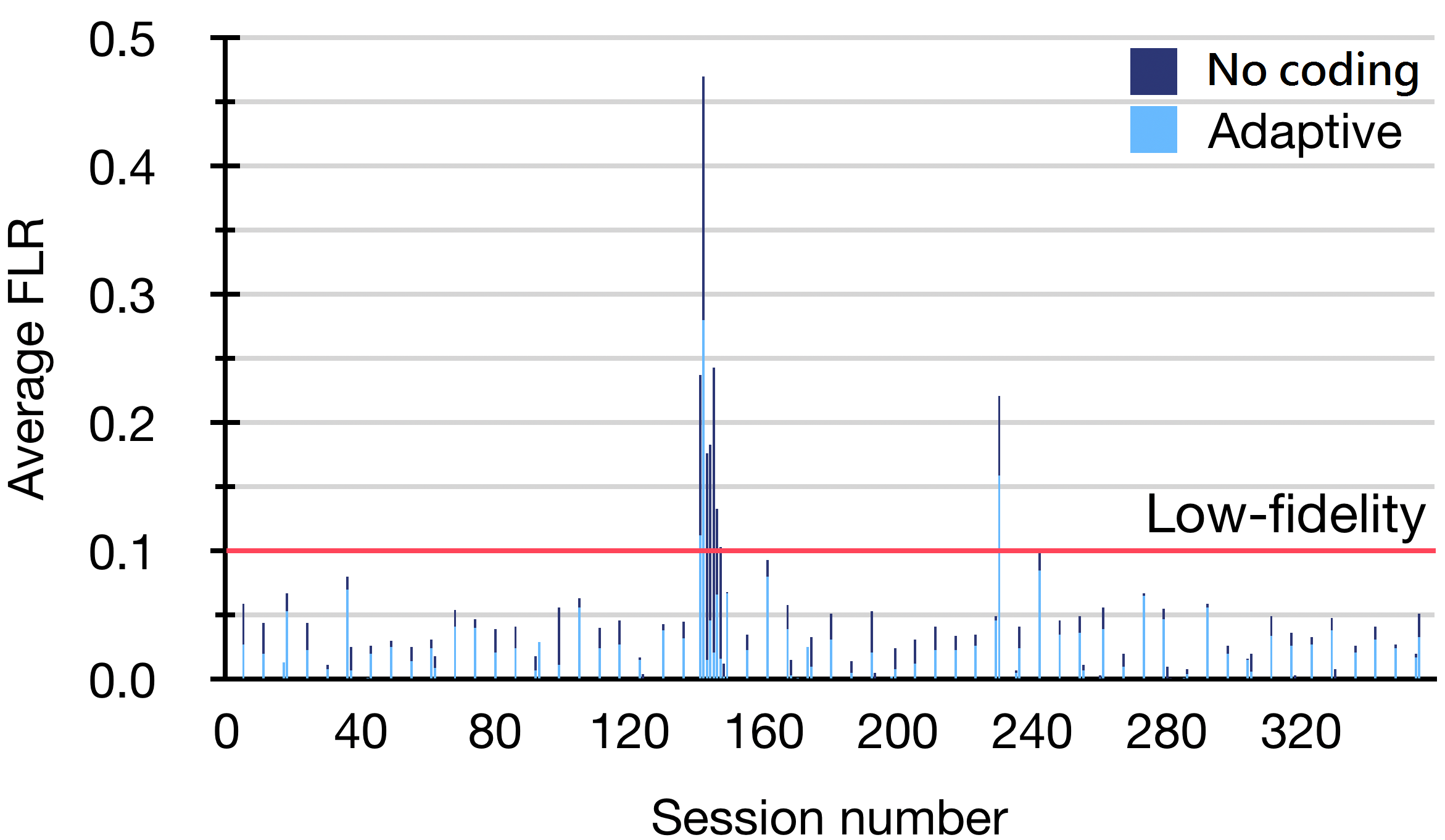}
\caption{Average FLRs for adaptive FEC over time for 40\%-capacity offered load}  \label{figureBarChart}
\end{figure}

\begin{figure}[t!]
\centering
\includegraphics[width =\linewidth, bb=0 0 1300 600]{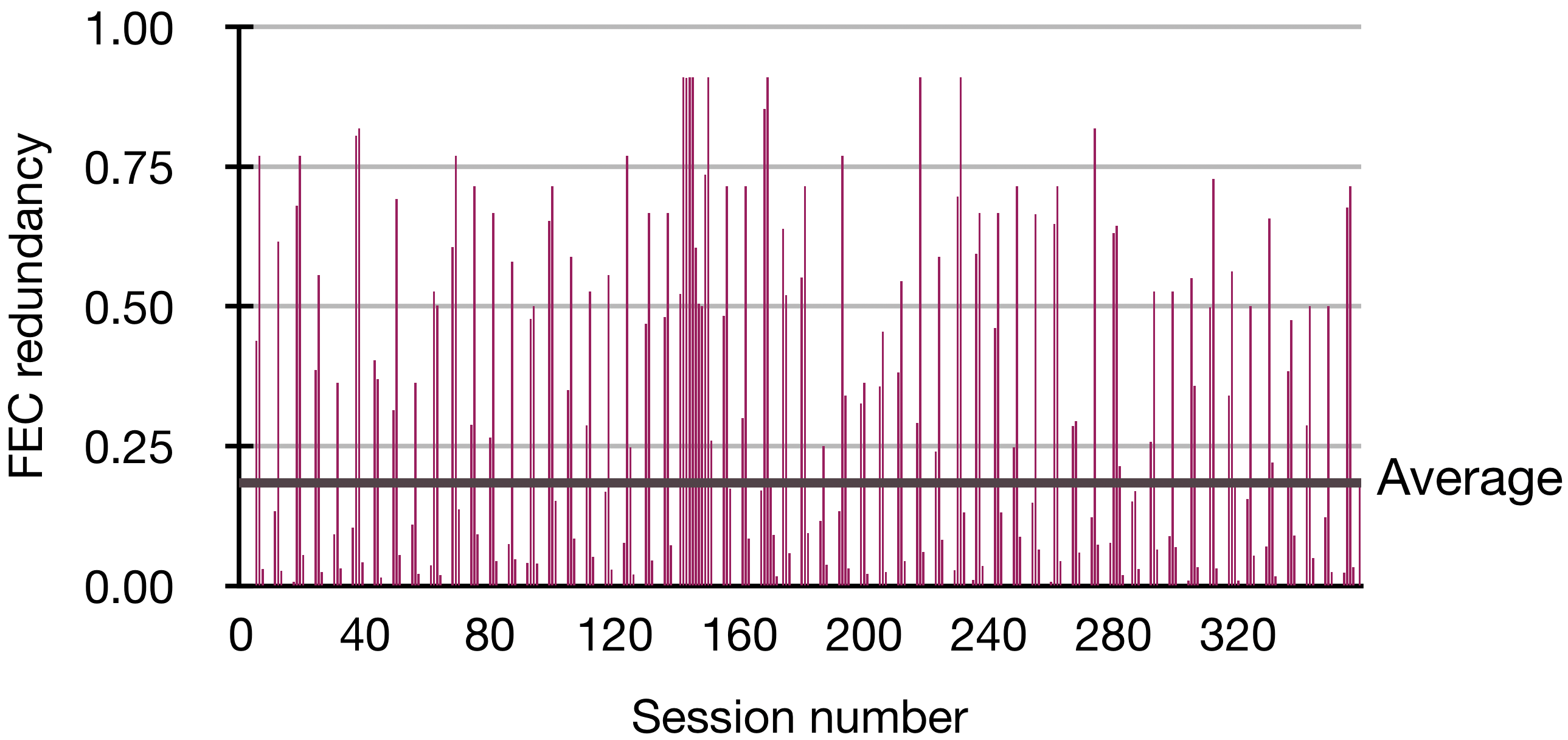}
\caption{FEC redundancy for 40\%-capacity offered load}\label{figureBarChartRedundancy}
\end{figure}
For the case where the offered load equals 40\% of the capacity, we display the empirical distribution of burst lengths in Fig.~\ref{histogram_wifi}. We observe a mean of burst lengths of 4.32 and standard deviation of 6.64. We also show in Fig.~\ref{figureBarChart} the variation of average FLRs for our adaptive streaming scheme and UDP across the $360$ sessions. It can be seen from Fig.~\ref{figureBarChart} that for more than 1/4 of the sessions that experience packet loss, our adaptive scheme achieves less than half of the no-coding loss rate. In addition, we display the variation of the FEC redundancy (i.e., one minus coding rate) for our adaptive scheme in Fig.~\ref{figureBarChartRedundancy}, which demonstrates how quickly it reacts to erasures.


\subsubsection{Experimental PESQ Scores} \label{subsubsecPESQ_Experiment}
In the previous subsection, we presented our experimental results in terms of FLRs for compressed multimedia frames where the duration and bit rate for each compressed frame are 10~ms and 240~kbit/s respectively. In this subsection, we follow the audio and codec settings as described in Section~\ref{subsubsecPESQ} (cf.\ Fig.~\ref{figureCodec}) and use the network-adaptive, uncoded or best non-adaptive scheme to transmit the compressed multimedia through the Wi-Fi network subject to Iperf UDP cross traffic as described in the previous subsection.

\begin{figure*}[t!]
\begin{minipage}[b]{.45\linewidth}
\centering \includegraphics[width = 0.9\linewidth]{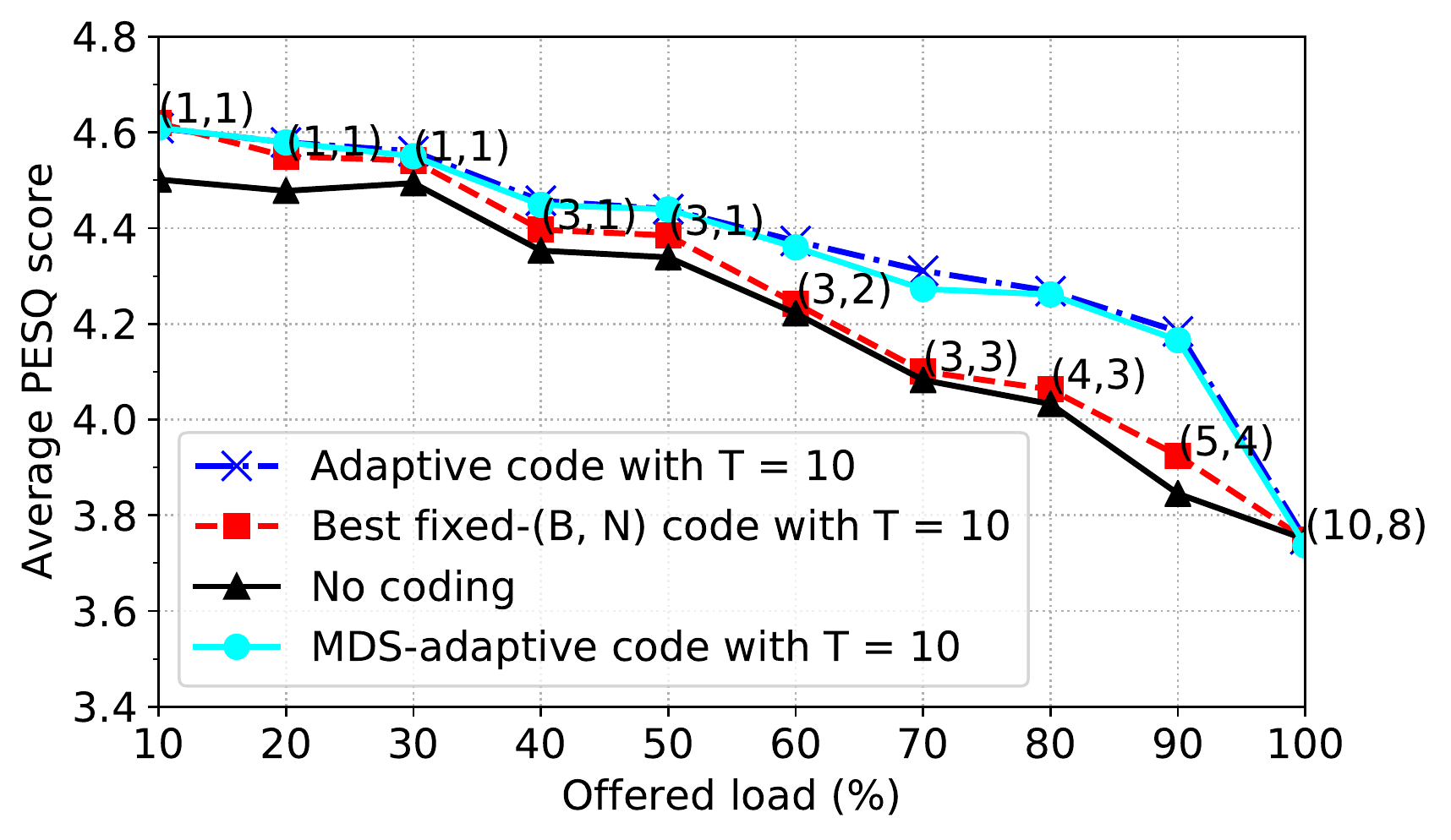}
\caption{Average PESQ score subject to Iperf traffic}
\label{figureLossRatePESQ}
\end{minipage}%
\hfill
\begin{minipage}[b]{.45\linewidth}
\centering \includegraphics[width =0.9 \linewidth]{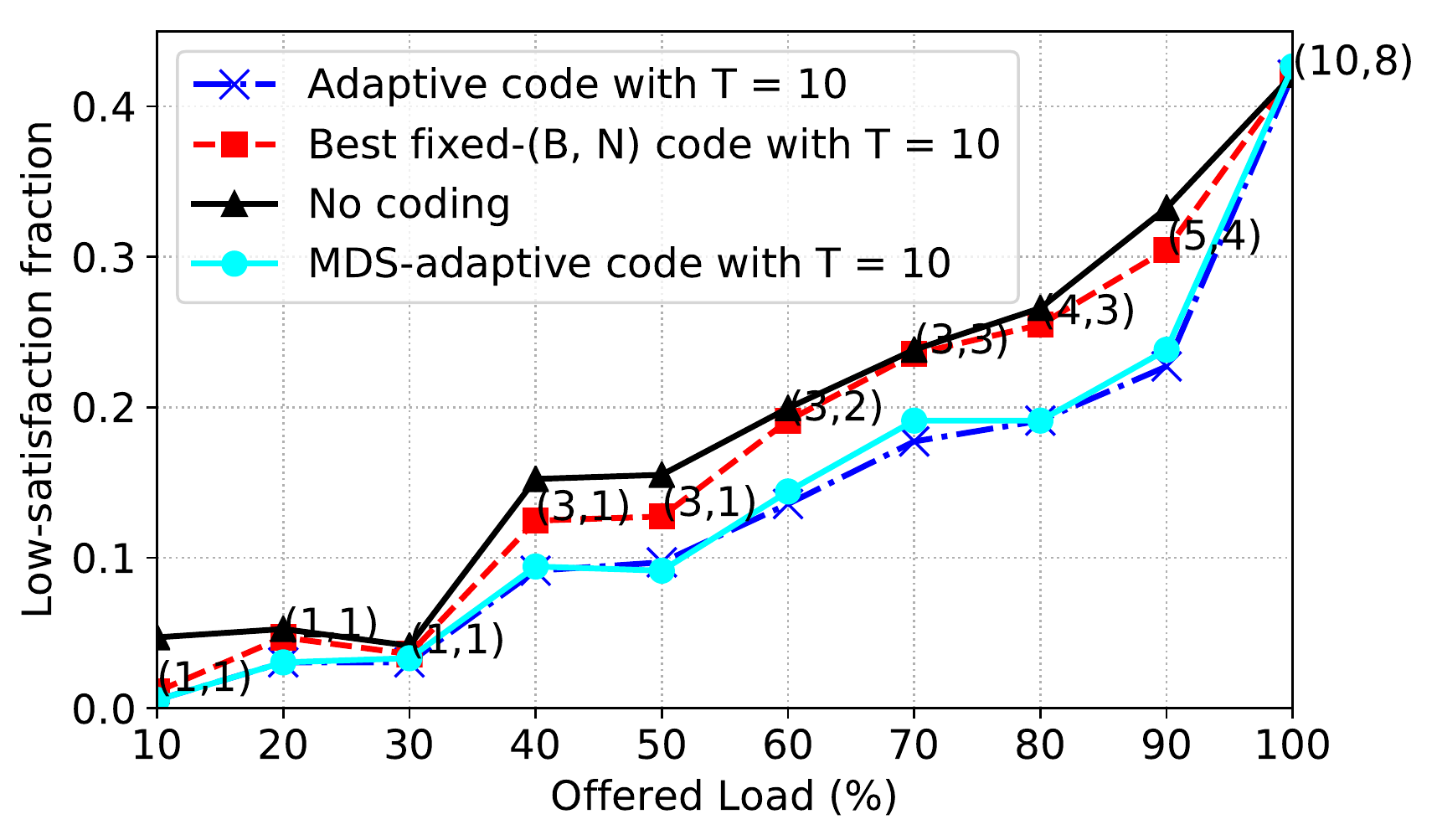}
\caption{Low-satisfaction fraction subject to Iperf traffic}
\label{figureLowFidelityPESQ}
\end{minipage}
\end{figure*}
For each 10-second session, a PESQ score is computed between the original uncompressed WAV audio and the recovered WAV audio for our all schemes. The average PESQ scores over 360 sessions for our network-adaptive streaming scheme is plotted against the percentage of the network capacity occupied by Iperf traffic in Fig.~\ref{figureLossRatePESQ}.

In addition, we use the recorded packet loss traces to simulate the average PESQ scores for the uncoded scheme and the best non-adaptive streaming code~$\mathcal{C}_{T, B, N}$ whose coding rate does not exceed the average coding rate of the network-adaptive scheme. The average PESQ scores for the uncoded scheme and the best non-adaptive streaming code with parameters $(B,N)$ are also plotted in Fig.~\ref{figureLossRatePESQ}.

We can see from Fig.~\ref{figureLossRatePESQ} that our adaptive streaming scheme achieves a higher average PESQ score than uncoded and non-adaptive streaming schemes. In addition, Fig.~\ref{figureLowFidelityPESQ} shows that our adaptive streaming scheme significantly reduces the fraction of low-satisfaction sessions (with PESQ score lower than 3.8) compared to the uncoded and non-adaptive schemes. As we mentioned earlier, the average of PESQ scores of affected sessions may better show the enhancement in the performance in extreme situations, hence Fig.~\ref{figurePESQAffectedWiFi} shows the average PESQ scores of affected sessions is higher for our scheme compared to uncoded and non-adaptive streaming schemes.

\begin{figure}[t!]
\centering
\includegraphics[width=0.9\linewidth]{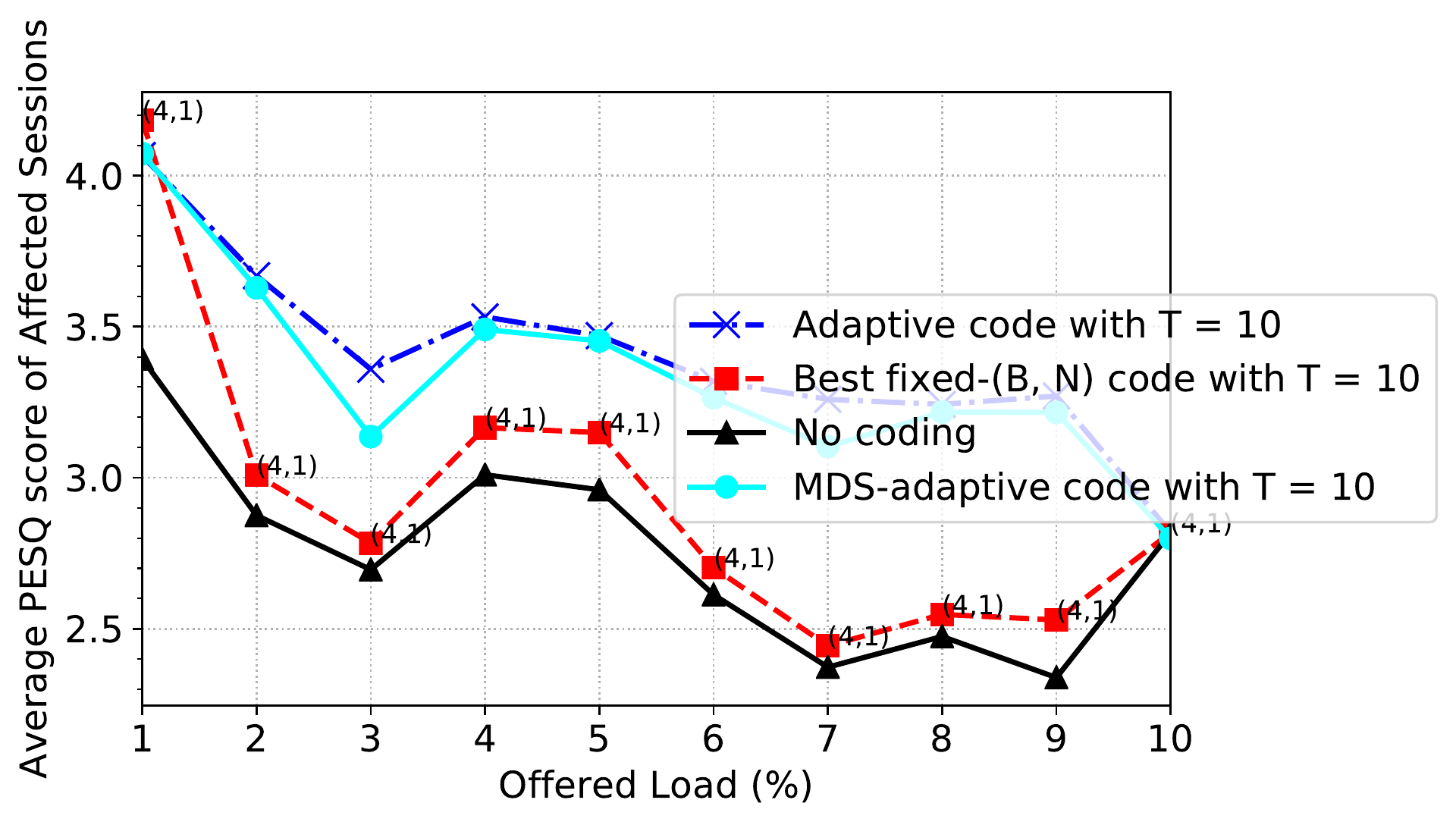}
\caption{Average PESQ of affected sessions subject to Iperf traffic} \label{figurePESQAffectedWiFi}
\end{figure}

\begin{figure}[t!]
\centering
\includegraphics[width = .9\linewidth]{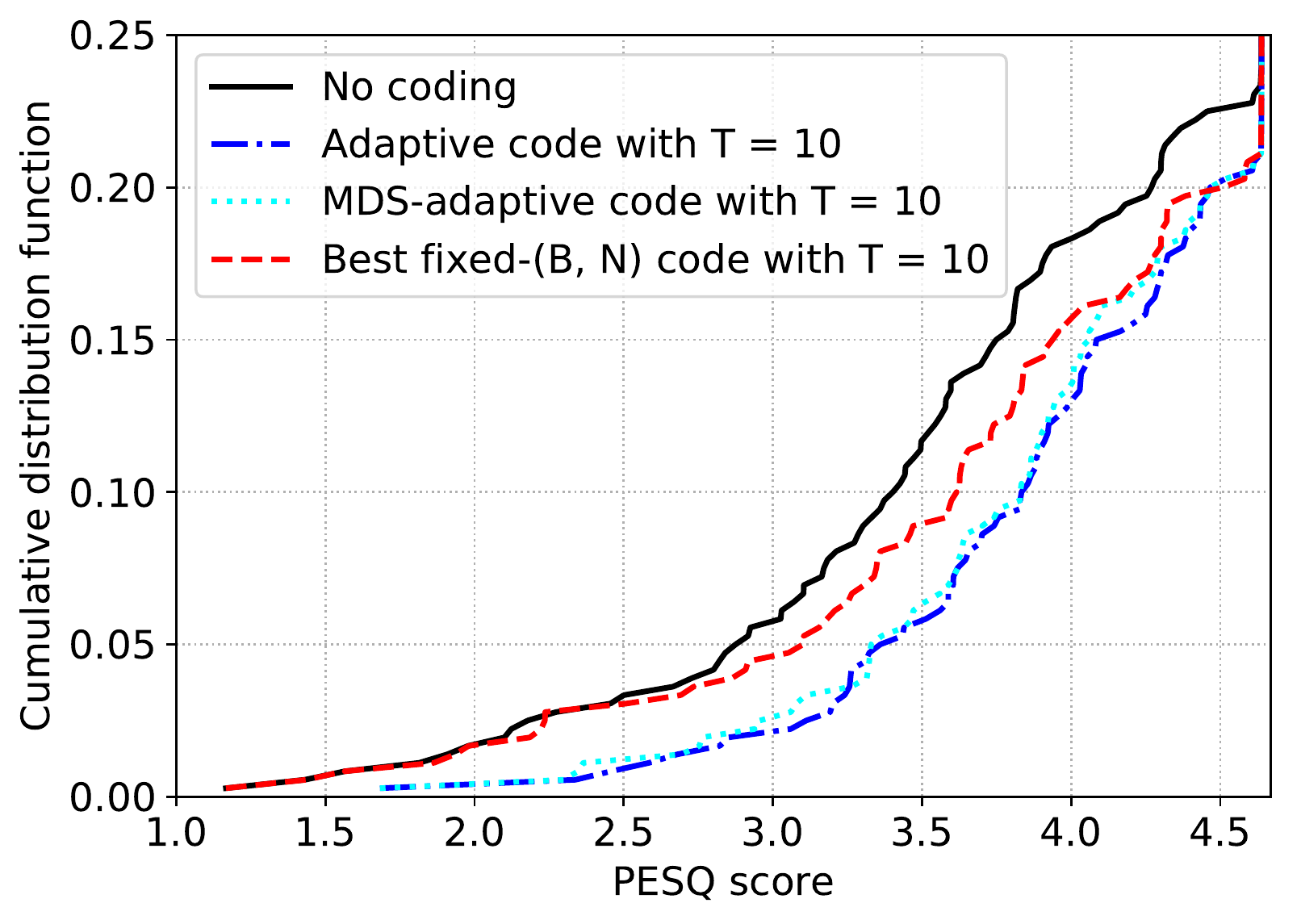}
\caption{The CDF of PESQ score for adaptive FEC, MDS-adaptive FEC and non-adaptive FEC for 40\%-capacity offered load}  \label{figureBarChartPESQ}
\end{figure}
We plot in Fig.~\ref{figureBarChartPESQ} the CDF of PESQ score (i.e., the fraction of the 360 sessions whose scores are less than a certain PESQ score) for each of the following schemes under 40\%-capacity offered load: Our adaptive streaming scheme, the best non-adaptive streaming code and the uncoded scheme. It can be seen from Fig.~\ref{figureBarChartPESQ} that our adaptive scheme provides the best user experience compared to the uncoded and non-adaptive schemes.

\subsubsection{Optimal Streaming Codes vs.\ MDS-Based Streaming Codes}  \label{subsubsecPESQ_Experiment_MDS}
\begin{figure}[t!]
\centering
\includegraphics[width=.9\linewidth]{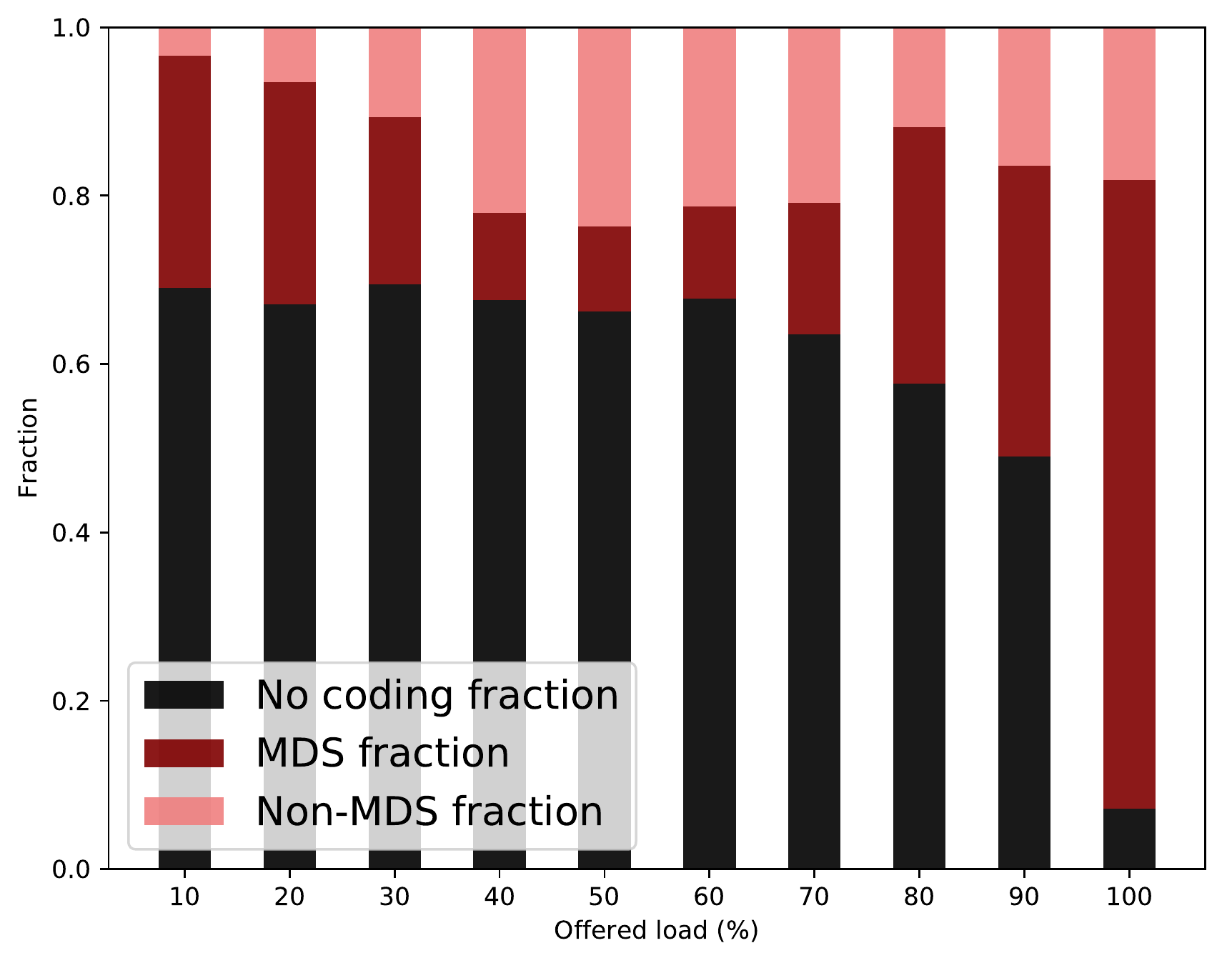}
\caption{Fraction of time where non-MDS codes are used.} \label{figure1WifiFraction}
\end{figure}

We plot in Fig.s~\ref{figureLossRate}, \ref{figureLowFidelity}, \ref{figureLossRatePESQ}, \ref{figureLowFidelityPESQ} and~\ref{figureBarChartPESQ} the respective average FLR, low-fidelity fraction, average PESQ score, low-satisfaction fraction and the PESQ CDF under 40\%-capacity offered load for the MDS-adaptive streaming scheme as described in Section~\ref{subsubsectionMDSstreaming}, which show that our network-adaptive scheme is marginally better than the MDS-adaptive scheme. This can be explained by Fig.~\ref{figure1WifiFraction} which shows that the fraction of non-MDS codes chosen by the adaptive algorithm used by our adaptive scheme is less than 25\% for all offered loads, implying that the optimal streaming codes chosen by our adaptive scheme are non-MDS less than 25\% of the time.

\begin{table}{
\centering
\begin{subtable}{0.5\textwidth}
\begin{tabular}{|c|*{4}{c|}}\hline
\textit{strategy} & \textit{FEC redundancy} & \textit{average FLR}& \textit{low-fi fraction} \\\hline
network-adaptive & {19.03}\% & 0.01190 & 0.03047  \\\hline
MDS-adaptive & 20.17\% & 0.01757 & 0.03878 \\\hline
no coding & 0\% & 0.03489 & 0.01856 \\\hline
\end{tabular}
\caption{$T=10$}
\vspace{0.05 in}
\label{table3a}
\end{subtable}
\begin{subtable}{0.5\textwidth}
\begin{tabular}{|c|*{4}{c|}}\hline
network-adaptive & \makebox[\widthof{\textit{FEC redundancy}}]{16.6\%} & \makebox[\widthof{\textit{average FLR}}]{0.02586} & \makebox[\widthof{\textit{low-fi fraction}}]{0.07202}  \\\hline
MDS-adaptive & 17.2\% & 0.02617 & 0.07756 \\\hline
no coding & 0\% & 0.04003 & 0.17452 \\\hline
\end{tabular}
\caption{$T=9$}
\vspace{0.05 in}
\label{table3b}
\end{subtable}
\begin{subtable}{0.5\textwidth}
\begin{tabular}{|c|*{4}{c|}}\hline
network-adaptive & \makebox[\widthof{\textit{FEC redundancy}}]{16.10\%} & \makebox[\widthof{\textit{average FLR}}]{0.00486} & \makebox[\widthof{\textit{low-fi fraction}}]{0.00831}  \\\hline
MDS-adaptive & 17.24\% & 0.01040 & 0.01108 \\\hline
no coding & 0\% & 0.02486 & 0.13573 \\\hline
\end{tabular}
\caption{$T=11$}
\label{table3c}
\end{subtable}
\caption{Performance of different streaming strategies for 30\%-capacity offered load}
\label{table3}
}
\end{table}
Since it is unclear from Fig.~\ref{figureLossRate}, \ref{figureLowFidelity}, \ref{figureLossRatePESQ} and~\ref{figureLowFidelityPESQ} that our network-adaptive streaming scheme can significantly outperform the MDS-adaptive streaming scheme in real-world networks, we use the recorded packet loss traces obtained from a repeated experiment with 30\%-capacity offered load to compare the network-adaptive streaming scheme with the MDS-adaptive streaming scheme for $T=10$.

Table~\ref{table3a} shows that although the two schemes have similar rates, the network-adaptive streaming scheme achieves around 80\% of the average FLR and 80\% of the low-fidelity sessions achieved by the MDS-adaptive streaming scheme, which implies that our constructed optimal streaming codes outperforms traditional MDS-based streaming codes in real-world networks.

The reduction in FLR is not surprising because our optimal streaming codes treat burst erasures and arbitrary erasures differently while MDS-based codes do not differentiate them. {Even if $T$ slightly deviates from~$10$, our experimental results displayed in Tables~\ref{table3b} and~\ref{table3c} show that the network-adaptive streaming scheme compared with the MDS-adaptive streaming scheme can achieve a considerable reduction in FLR.} In particular, for $T=11$, although the two adaptive schemes have similar rates, the network-adaptive streaming scheme achieves around 50\% of the average FLR and 75\% of the low-fidelity sessions achieved by the MDS-adaptive streaming scheme.

In our setting, due to the 150ms one-way delay constraint and the standard 10ms frame duration assumption, restricting T<=11 is a reasonable assumption. The search for efficient streaming codes over GF(256) or other practical fields for T>11 is an interesting direction for future research. 

\section{Conclusion and Future Work} 
\label{conclusion}
We have designed a network-adaptive FEC streaming scheme which consists of (i) a network-adaptive algorithm for estimating the coding parameters of streaming codes that correct both burst and arbitrary network packet losses, and (ii) an explicit construction of low-latency optimal streaming codes over GF(256) for $T\le 11$. {The computation bottleneck of our network-adaptive streaming scheme is bounded above by $O(T^3)$ complexity, where the bottleneck upper bound is due to the complexity of decoding a length-$(T+1)$ block code using Gauss-Jordan elimination~\cite{Farebrother:1988:LLS:59113}. More precisely, the computation bottleneck is close to $O(D^3)$ where~$D$ denotes the average number of lost packets in a sliding window of size~$T+1$, where the bottleneck is due to the average complexity of decoding the lost packets in a sliding window of size~$T+1$.} In simulated and real-world experiments, our results reveal that our adaptive streaming scheme significantly outperforms non-adaptive and uncoded ones in terms of FLRs and PESQ scores. We also highlight the advantage of using our network adaptive scheme compared to MDS-adaptive schemes. 

There are several interesting directions for future investigation: (i) Finding the largest~$T$ such that optimal streaming codes exist over GF(256) remains open. (ii)  Future work may explore the interplay between our adaptive streaming scheme which adjusts the coding rate in real time and existing congestion control algorithms that adjust the sizes of streaming messages in real time. (iii) Also, machine learning techniques could be used to develop new network-adaptive algorithms. 



%
%
%
%




\ifCLASSOPTIONcaptionsoff
  \newpage
\fi




\end{document}